\documentclass[10pt,aps,prd,twocolumn,superscriptaddress,nofootinbib]{revtex4-2}
\usepackage{amsmath, amssymb, bm}  
\usepackage{graphicx}  
\usepackage{tensor} 
\usepackage{siunitx}
\usepackage[colorlinks=true,linkcolor=blue,citecolor=blue,urlcolor=blue]{hyperref} 
\usepackage{lineno}
\usepackage{orcidlink}
\begin{document}

\title{\texorpdfstring{GFH-v2 Pipeline for Searches of Long-Transient Gravitational Waves\\
from Newborn Magnetars}{GFH-v2 Pipeline for Searches of Long-Transient Gravitational Waves from Newborn Magnetars}}

\author{Sandhya Sajith Menon\,\orcidlink{0009-0008-4985-1320}}
\affiliation{Physics Department, Ariel University, Ariel, Israel}
\affiliation{Istituto Nazionale di Fisica Nucleare-Roma 1, Piazzale Aldo Moro 2, I-00185, Rome, Italy}
\affiliation{Università di Roma La Sapienza, I-00185 Rome, Italy}

\author{Lorenzo Pierini\,\orcidlink{0000-0003-0945-2196}}
\affiliation{Istituto Nazionale di Fisica Nucleare-Roma 1, Piazzale Aldo Moro 2, I-00185, Rome, Italy}

\author{Pia Astone\,\orcidlink{0000-0003-4981-4120}}
\affiliation{Istituto Nazionale di Fisica Nucleare-Roma 1, Piazzale Aldo Moro 2, I-00185, Rome, Italy}

\author{Cristiano Palomba\,\orcidlink{0000-0002-4450-9883}}
\affiliation{Istituto Nazionale di Fisica Nucleare-Roma 1, Piazzale Aldo Moro 2, I-00185, Rome, Italy}

\author{Lorenzo Silvestri\,\orcidlink{0009-0008-5207-661X}}
\affiliation{Università di Roma La Sapienza, I-00185 Rome, Italy}
\affiliation{INFN-CNAF - Bologna, Viale Carlo Berti Pichat, 6/2, 40127 Bologna BO, Italy}

\author{Sabrina \surname{D'Antonio}\,\orcidlink{0000-0003-0898-6030}}
\affiliation{Istituto Nazionale di Fisica Nucleare-Roma 1, Piazzale Aldo Moro 2, I-00185, Rome, Italy}

\author{Simone Dall’Osso\,\orcidlink{0000-0003-4366-8265}}
\affiliation{Dipartimento di Fisica e Astronomia, Alma Mater Studiorum - Università di Bologna, I-40127 Bologna, Italy}

\author{Francesco \surname{Safai Tehrani}\,\orcidlink{0000-0001-7796-0120}}
\affiliation{Istituto Nazionale di Fisica Nucleare-Roma 1, Piazzale Aldo Moro 2, I-00185, Rome, Italy}

\author{Stefano \surname{Dal Pra}\,\orcidlink{0000-0002-1057-2307}}
\affiliation{INFN-CNAF - Bologna, Viale Carlo Berti Pichat, 6/2, 40127 Bologna BO, Italy}

\author{Gaetano Dinatale}
\affiliation{Università di Roma La Sapienza, I-00185 Rome, Italy}

\author{Sergio \surname{Frasca}\,\orcidlink{0000-0002-0898-6703}}
\affiliation{Istituto Nazionale di Fisica Nucleare-Roma 1, Piazzale Aldo Moro 2, I-00185, Rome, Italy}

\author{Dafne Guetta\,\orcidlink{0000-0002-7349-1109}}
\affiliation{Physics Department, Ariel University, Ariel, Israel}

\author{Paola Leaci\,\orcidlink{0000-0002-3997-5046}}
\affiliation{Istituto Nazionale di Fisica Nucleare-Roma 1, Piazzale Aldo Moro 2, I-00185, Rome, Italy}
\affiliation{Università di Roma La Sapienza, I-00185 Rome, Italy}

\author{Alessio Orlandi}
\affiliation{Università di Roma La Sapienza, I-00185 Rome, Italy}
\date{\today}

%\linenumbers
\begin{abstract}
    This paper presents an enhanced methodology for searching long transient gravitational waves associated with a newborn magnetar, with particular focus on the regime in which the early spin-down is dominated by gravitational-wave emission. The analysis is performed using a strongly improved version of the generalized Frequency Hough Transform algorithm, called GFH-v2. We describe the main developments introduced relative to the original implementation and outline the optimized parameter-space selection used in the search. We then compute the theoretical sensitivity of the method and compare it with an empirical sensitivity estimate obtained by injecting simulated signals into LIGO-Virgo-KAGRA O4a data. The updated framework achieves improved sensitivity and computational performance. These results provide a robust basis for future directed searches for long-transient gravitational-wave signals from core-collapse supernovae and other transient events in current and upcoming observing runs.
\end{abstract}

\maketitle

\section{Introduction}

Continuous gravitational waves (CWs) are long-duration, nearly monochromatic signals typically emitted by non-axisymmetric rotating neutron stars \cite{2015PASA3234L,2023LRR263R,2023APh15302880W}. Although still undetected, these signals will provide a powerful probe of neutron star properties, including their internal structure and spin evolution \cite{2011GReGr43409A}. Searches for CWs have so far mainly focused on isolated or binary neutron stars, either known from electromagnetic observations or searched in all-sky surveys \cite{2023LRR263R,2023APh15302880W,2022ApJ9351A}. However, within the broad category of CWs lies a particularly intriguing subclass: long-transient gravitational waves (tCWs, a term first introduced by Prix et al. \cite{2011PhRvD84b3007P} to describe signals of intermediate duration) \cite{2015PASA3234L,2023LRR263R,2011PhRvD83j4014C}. These are signals with durations ranging from thousands of seconds to days, characterized by non-linear frequency evolution due to the intense rotational energy loss of the source.

tCWs represent a promising yet challenging frontier in gravitational wave (GW) observations. Among the potential sources of such signals are newborn magnetars: rapidly rotating, highly magnetized neutron stars formed in the aftermath of catastrophic events such as core-collapse supernovae (CCSNe) or binary neutron star (BNS) mergers \cite{2001A&A367525P,2002PhRvD66h4025C,2009ApJ701171C,2021MNRAS5024680S,2022ASSL465245D}. In newborn magnetars, gravitational wave emission may arise from non-axisymmetric deformations induced by strong internal magnetic fields, residual asymmetries from the formation process, or other mechanisms such as bar-mode instabilities or fallback accretion.

Detecting tCWs offers a unique opportunity to probe extreme physics that is otherwise inaccessible: the equation of state (EoS) of ultra-dense matter, the internal magnetic field configuration, and the dynamics of the newly formed remnant \cite{2019EPJA5580R,2022NatRP4237Y}. Additionally, this information could give us significant insight on the nature of the progenitor and the physical mechanisms during and after the merger or the core-collapse \cite{2019ApJ875160A}.

However, these signals are difficult to detect due to their weak amplitude and broad parameter space. Hierarchical search strategies, particularly those employing semi-coherent techniques, are essential to maintain sensitivity while keeping computational costs tractable. Semi-coherent methods split the dataset into shorter segments, analyze each coherently, and then combine the results incoherently. This approach reduces computational cost while retaining most of the sensitivity of a fully coherent search. Several analysis methods have been developed
to search for long-transient continuous gravitational waves
from neutron stars, including transient extensions of the
$\mathcal{F}$-statistic and other approaches
designed to account for signal durations ranging from
hours to days \cite{2011PhRvD84b3007P,2019PhRvD.100b4034B,2018CQGra35t5003K,2019PhRvD99j4067O,2019PhRvD99l3003S,2019PhRvD.100f4058K,2016PhRvD93h4024K,2024PhRvD.109l3516A,2025PhRvD.111d3019A}. One such approach is the Generalized Frequency Hough (GFH) pipeline \cite{2018PhRvD98j2004M}, which was previously applied to the post-merger remnant of GW170817, though no signal was observed at the time, as the source distance was much larger than the potential reach of the search \cite{2019ApJ875160A}.

Building on that foundation, in this paper we introduce an improved version of the method, which we call GFH-v2. It incorporates significant enhancements of the algorithm, resulting in higher sensitivity and better management of computational resources. As a consequence, it offers more promising detection prospects for the upcoming O5 science run of the LIGO-Virgo-KAGRA \cite{2020LRR233A} detectors and future third-generation detectors, like ET \cite{Maggiore2020_ET} and Cosmic Explorer \cite{Evans2021_CE}.

The paper is structured as follows.
In Sec. \ref{sec:magne} we briefly describe magnetars, the relevant equations describing their rotational evolution and the main characteristics of the associated gravitational wave emission. Sec. \ref{sec:gfh} provides the basic concepts of the original GFH algorithm, while the new version, GFH-v2 and its methodological improvements are discussed in detail in Sec. \ref{sec:gfhv2}. Within this section, we discuss in detail the astrophysically motivated search parameter space, including the choice of coherence time and observation-time window, as well as the technical enhancements to the analysis pipeline.

Sec.~\ref{sec:theory_sens} deals with the semi-theoretical estimation of the method’s sensitivity, based on the assumption of Gaussian noise. In Sec.~\ref{sec:exp_sens}, we compare the theoretical results with an experimental computation, obtained by injecting simulated signals into data from the first part of the fourth LIGO-Virgo-KAGRA observing run. This section also includes detailed discussions on data characteristics, the injection setup, candidate selection, and efficiency analysis.

Finally, Sec.~\ref{sec:concl} presents conclusions and future prospects of this work. Additional technical details, including the derivation of the theoretical sensitivity formula and the procedure adopted for the selection of the critical ratio threshold, are provided in Appendices~\ref{app:theosens} and~\ref{app:crthr}, respectively.

\section{Newborn Magnetars and associated GW signal}
\label{sec:magne}

Newborn magnetars, a type of highly magnetized neutron star, are thought to form in the aftermath of CCSNe explosions or in BNS mergers. These compact objects are characterized by extremely strong surface magnetic fields, typically of the order \(10^{14}-10^{15} \, \mathrm{G}\), and rapid initial spin periods of a few milliseconds ($P_{\rm 0} \sim 1-10$ ms). Magnetars are typically born from progenitor stars with masses in the range of \(10\) to \(30 \, M_\odot\)\cite{2022ASSL465245D} and their formation rate in the galaxy is estimated to be $2.3$-$20$ per kyr, corresponding to roughly $10$-$15\%$ of all CCSNe \cite{2019MNRAS4871426B}

At the time of their birth, magnetars are expected to possess asymmetries resulting in a highly variable mass quadrupole moment, leading to the emission of gravitational waves \cite{2009MNRAS3981869D,2012ApJ76163P,dallo15,2018MNRAS4801353D,2020MNRAS4944838L}. Mechanisms driving these asymmetries include strong interior magnetic fields, a secular bar-mode instability~\cite{1995ApJ442259L},
formation of accretion-induced mountains on the surface~\cite{2000MNRAS.319902U,2008MNRAS.385531H} etc., which generate long-lived non-axisymmetric deformations, which are necessary for GW emission. GW emission, with power \( \dot{E}_{\text{GW}} \), happens at the expense of the star's rotation frequency according to \cite{Maggiore:2007ulw}
\begin{equation}
    \dot{E}_{\text{GW}} = \frac{32}{5} \frac{G \epsilon^2 I^2 \Omega^6}{c^5},
    \label{eq:Edot}
\end{equation}
where $\epsilon = (I_{xx}-I_{yy})/I_{zz}$ is the ellipticity, defined in terms of principal moment of inertia, $I\equiv I_{zz}$ is the moment of inertia w.r.t the rotation axis, and \( \Omega \) is the angular frequency. Throughout this work we assume that the ellipticity remains constant over the signal duration. A time-dependent ellipticity would modify both the GW amplitude and the spin-down evolution of the signal and is beyond the scope of this study. From Eq. \ref{eq:Edot}, the angular frequency decreases as:
\begin{equation}
    \dot{\Omega} = -\frac{32}{5} \frac{G \epsilon^2 I \Omega^5}{c^5}.
    \label{eq:ang_velocity}
\end{equation}
These equations highlight how magnetars can efficiently lose rotational energy via GW emission, especially in the early stages of their evolution. In this work, we focus on the simplified case in which the early spin-down of the source is dominated by GW emission. This represents a limiting scenario where GW emission is assumed to be stronger than magnetic dipole radiation. In reality, newborn magnetars may experience a combination of spin-down mechanisms. However, adopting the GW-dominated case allows us to work with a well-defined signal model that is suitable for developing and testing the analysis pipeline. More general scenarios including additional spin-down contributions will be explored in future work.

The GW signal emitted by a rotating rigid neutron star has an amplitude proportional to the square of the GW frequency \cite{1998PhRvD58f3001J}. For systems where the spin frequency evolves with time, such as newly born magnetar, the amplitude becomes time dependent through the instantaneous GW frequency $f_{\rm gw}(t)$:
\begin{equation}
    h_0(t) = \frac{4 \pi^2 G}{c^4} \frac{I \epsilon}{d} f_{\rm gw}^2(t),
    \label{eq:amplitude}
\end{equation}

For a rigid non-precessing neutron star rotating about a principal axis, the GW emission occurs at twice the rotational frequency\footnote{This relation holds for non-precessing NSs rotating about a principal axis. Other mechanisms, such as $r$-modes or precession, may produce GW signals at different frequencies and with more complex evolution.}~\cite{Maggiore:2007ulw}:

\begin{equation}
    f_{\rm gw} = 2 f_{\rm rot}\, =\Omega/\pi,
    \label{eq:fgw_frot}
\end{equation}

The rotational frequency evolution of the magnetar can be described, in general, by a power law spin-down model,

\begin{equation}
    \dot{f}_{\rm rot} \propto - f_{\rm rot}^n,
    \label{eq:spindown}
\end{equation}
where n is the braking index which depends on the physical mechanism responsible for the spindown (e.g., magnetic dipole radiation, GW emission, or particle winds)~\cite{2015PhRvD91f3007H}. 

The corresponding evolution for the GW frequency can then be written as
\begin{equation}
    \dot{f}_{\rm gw} = -k f_{\rm gw}^n,
    \label{eq:spindown_gw}
\end{equation}

Integrating Equation~\ref{eq:spindown_gw} yields the frequency evolution:
\begin{equation}
    f_{\rm gw}(t) = \frac{f_0}{ \left(1 + \frac{t - t_0}{\tau} \right)^{1/{(n-1)}} },
    \label{eq:frequency}
\end{equation}
where $f_0$ is the initial frequency at reference time $t_0$ and the characteristic timescale $\tau$ is:
\begin{equation}
    \tau = \frac{1}{(n-1) k f_0^{n-1}}.
\end{equation}

The proportionality constant $k$ depends on the underlying emission process. In this work, we focus on the GW-driven spin-down case with $n=5$~\cite{2008PhRvD78d4031K,Shapiro:1983du}, assuming rotation about a principal axis of inertia for which the proportionality constant is
\begin{equation}
k = \frac{32 \pi^4 G}{5 c^5} \epsilon^2 I.
\end{equation}

Substituting Equation~\ref{eq:frequency} into Equation~\ref{eq:amplitude}, the amplitude becomes:
\begin{equation}
    h_0(t) = \frac{4 \pi^2 G}{c^4} \frac{I \epsilon}{d} f_0^2 \left(1 + \frac{t - t_0}{\tau} \right)^{-1/2}.
    \label{eq:amplitude_evolved}
\end{equation}

\begin{figure}[h]
    \centering
    \includegraphics[width=0.47\textwidth]{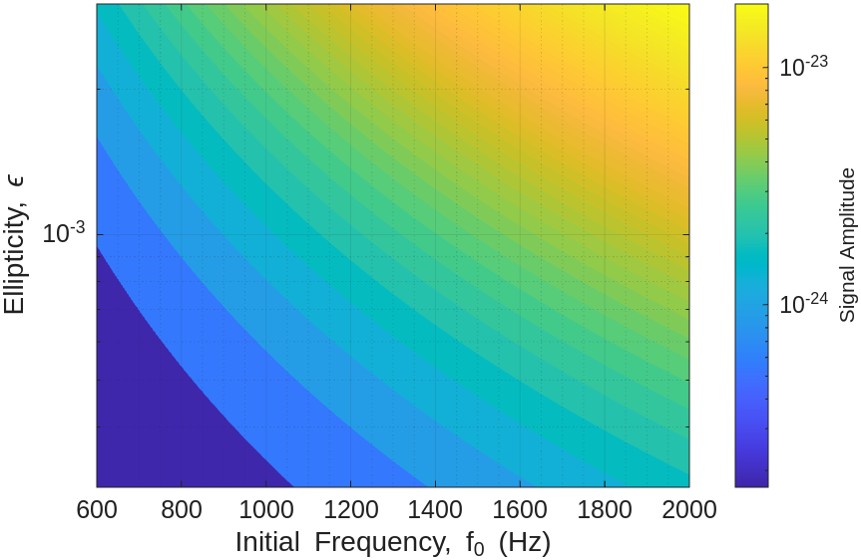}
    \caption{Gravitational-wave strain amplitude $h_0(t=t_0)$ as a function of initial spin frequency $f_{\rm gw}(t_0)\equiv f_0$ and ellipticity $\epsilon$. The source distance is fixed to 1 Mpc for illustration purposes. The moment of inertia is computed using Eq.~\ref{eq:Moment_I} for $M=1.4\,M_\odot$ and $R=12\,\mathrm{km}$.}
    \label{fig:amplitude_vs_f0_ellipticity}
\end{figure}

Figure~\ref{fig:amplitude_vs_f0_ellipticity} illustrates the dependence of signal amplitude on initial spin frequency and ellipticity for a range of plausible values.

The moment of inertia $I$ can be approximated as a function of the compactness $\beta$ using the empirical formula \cite{LATTIMER2016127}:
\begin{equation}
    I \approx M R^2 \left(0.247 + 0.642\beta + 0.466\beta^2 \right),
    \label{eq:Moment_I}
\end{equation}

where
\begin{equation}
    \beta = \frac{G M}{c^2 R}.
\end{equation}

This relation holds for $\beta > 0.1$, and assumes a neutron star with a maximum mass $M_{\rm max}$ (the maximum mass allowed by the equation of state) satisfying $M_{\rm max} \geq 1.97\, M_{\odot}$.

\section{generalized Frequency Hough}
\label{sec:gfh}
\begin{figure*}[ht]
    \centering
    \begin{minipage}{0.50\textwidth}
        \centering
        \includegraphics[width=\linewidth]{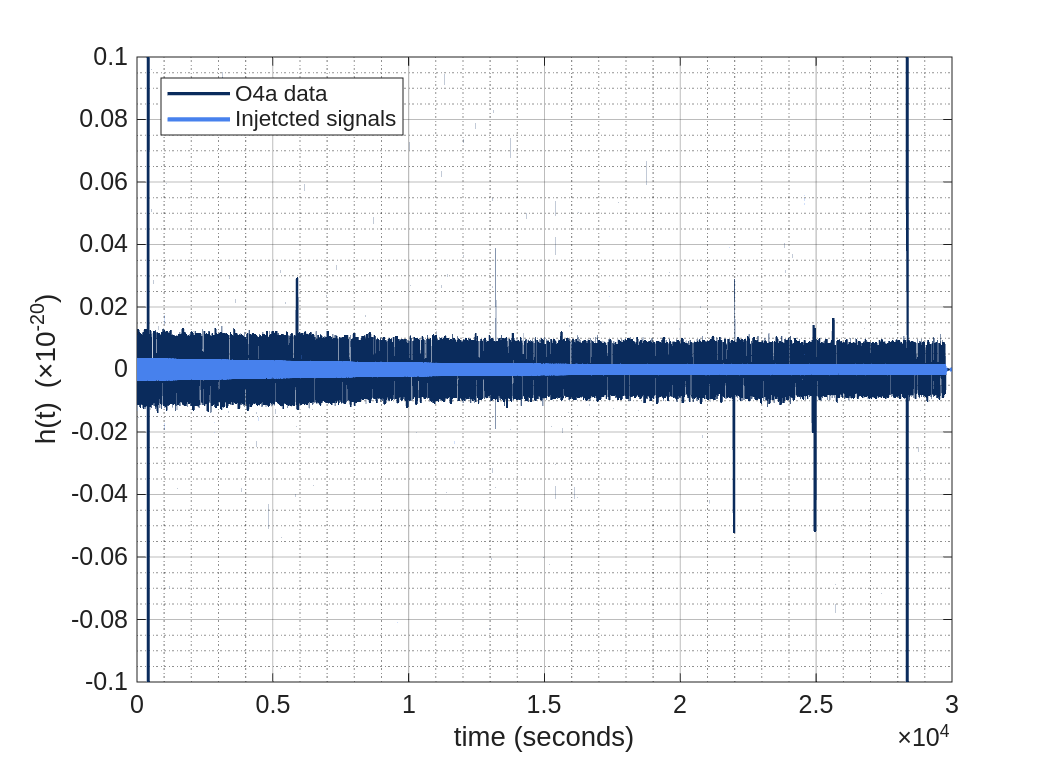}\\[4pt]
        \includegraphics[width=\linewidth]{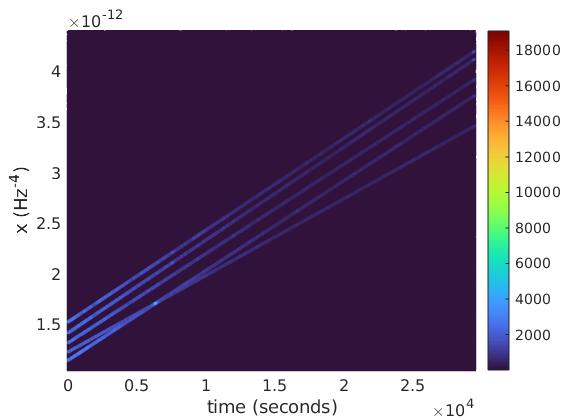}
    \end{minipage}
    \begin{minipage}{0.49\textwidth}
        \centering
        \includegraphics[width=\linewidth]{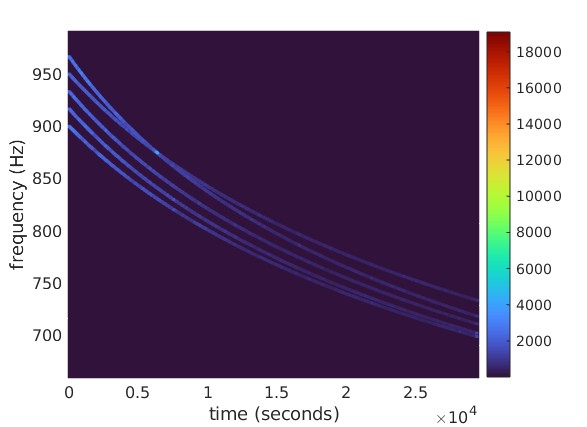}\\[4pt]
        \includegraphics[width=\linewidth]{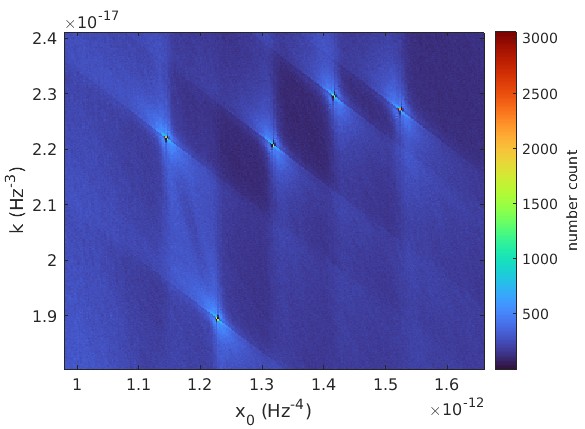}
    \end{minipage}
    \caption{
    Workflow of the GFH transform. 
    \textit{Top left:} Raw strain data with injected signals. 
    \textit{Top right:} Original time-frequency peakmap. 
    \textit{Bottom left:} Coordinate-transformed peakmap, where signal trajectories become linear; color indicates normalized power of the peaks (applies also to the original peakmap above). 
    \textit{Bottom right:} Resulting Hough map, corresponding to 5 simulated signals injected in O4a LIGO Livingston data~\cite{2025arXiv250818079T,GWOSC} with $f_0 = 900$-$1000$~Hz, $\epsilon = (2.5$-$3)\times10^{-3}$, at a distance of 0.1~Mpc and sky location of SN2023ixf.
    }
    \label{fig:gfh_pipeline}
\end{figure*}
The Generalized Frequency Hough (GFH) transform is a semi-coherent, hierarchical search algorithm designed to detect long-transient gravitational wave signals whose frequency evolves non-linearly over time \cite{2018PhRvD98j2004M}. It extends the original Frequency Hough (FH) method \cite{2004PhRvD70h2001K,2014PhRvD90d2002A}, which was primarily developed for detecting continuous wave signals with linearly evolving frequency.

The method starts from the \textit{Short Fourier Transform Databases} (SFDB)~\cite{2005CQGra22S1197A, Astone2025SFDB}, which contain sequences of short Fourier transforms (SFTs) generated from calibrated strain data~\cite{2020CQGra37v5008S,2022CQGra39d5006A}. The data are initially divided into overlapping segments of equal length, with a 50\% overlap. Each segment has duration $T_{\mathrm{FFT}}$ and is transformed using the Fast Fourier Transform (FFT) algorithm. The SFDB spans the full frequency band (0-2000 Hz), with FFTs typically covering timescales of the order of thousands of seconds. We first read the SFDB and extract the desired frequency band. The FFTs are then inverse-transformed back to a time series and subdivided into shorter segments with the desired FFT length for the analysis.

The construction of the \textit{Peakmap} proceeds by computing, for each of the $N$ FFTs, the ratio between the periodogram (i.e., the squared modulus of the FFT) and the corresponding average spectral estimate computed in the frequency domain~\cite{2014PhRvD90d2002A}. The most significant peaks in this ratio, that is, local maxima whose normalized power exceeds a predefined threshold, are identified and collected to form a time-frequency map of selected peaks. 

The original FH transform maps these peaks from the time-frequency plane of the detector into the frequency-spindown parameter space of the source, assuming a linear frequency evolution. In this case, the gravitational-wave frequency can be written as a Taylor expansion around the reference time $t_0$:
\begin{equation}
f(t) = f_0 + \dot{f}(t-t_0) ,
\label{eq:fh_freq_evol}
\end{equation}
where $f_0$ is the source frequency at $t_0$, and $\dot{f}$ is the first-order spindown parameter. In the FH framework, each peak $(t,f)$ in the detector’s peakmap is mapped into a line in the $(f_0, \dot{f})$ source plane:
\begin{equation}
\dot{f} = -\frac{f_0}{t-t_0} + \frac{f}{t-t_0}.
\label{eq:fh_mapping}
\end{equation}
Clusters of intersecting lines in this parameter space correspond to candidate CW signals.

However, this linear approximation breaks down for newborn magnetars or other transient sources with rapid frequency decay, where the frequency evolution instead follows a power law (see Eq.~\ref{eq:frequency}). 

The GFH transform overcomes this limitation by introducing a change of variables,
\begin{equation}
x = \frac{1}{f^{n-1}},
\quad
x_0 = \frac{1}{f_0^{n-1}},
\label{eq:x_transform}
\end{equation}
which linearizes the trajectory. In this $(t, x)$ space, the frequency evolution becomes a straight line,
\begin{equation}
x(t) = x_0 + (n-1)k(t-t_0).
\label{eq:gfh_line}
\end{equation}
Each peak $(t,x)$ in the transformed peakmap is then mapped into a line in the $(x_0, k)$ space:
\begin{equation}
k = -\frac{x_0}{(n-1)(t-t_0)} + \frac{x}{(n-1)(t-t_0)}.
\label{eq:gfh_mapping}
\end{equation}

As in the FH, true signals appear as clusters of intersecting lines in this parameter space, but now the method remains efficient even for rapidly evolving sources with power-law frequency decay.

An example of the GFH workflow is shown in Fig.~\ref{fig:gfh_pipeline}. The upper panels display the input strain and its corresponding original \textit{peakmap}, while the lower panels show the coordinate-transformed \textit{peakmap} and the resulting Hough map where the clusters indicate potential signal candidates. The clusters appear slanted because the reference time $t_0$ is set at the start of the dataset.

The Hough maps are built to cover a range of $(x_0, k)$ bins according to two discrete grids. The grid spacings $\delta x_0$ and $\delta k$ are chosen following the procedure described in \cite{2018PhRvD98j2004M}. Specifically, $\delta k$ is set such that the spindown remains constant when moving one frequency bin $\delta f = 1/T_{\rm FFT}$. The spacing $\delta x_0$ is chosen as the smallest step corresponding to the maximum frequency analyzed. With these choices, the grids uniformly cover the $(x_0, k)$ parameter space, ensuring that all signal trajectories are properly sampled in the Hough map.

To uniformly explore the parameter space, they are then partitioned into rectangles, where significant candidates are selected as excesses of counts and are then classified using a critical ratio statistic:
\begin{equation}
\mathrm{CR} = \frac{y - \mu}{\sigma},
\end{equation}
where $y$ is the number count in the outlier bin of the map, and $\mu$ and $\sigma$ are the mean value and standard deviation computed in the map, to estimate the noise floor. Candidates from different detectors are compared to identify coincident events, further reducing false alarms. Following \cite{2014PhRvD90d2002A}, candidates are considered coincident if they are close within a defined distance in the $(x_0, k)$ parameter space, computed as  
\begin{equation}
d_{\mathrm{coin}} = \sqrt{\left(\frac{x_{0,2} - x_{0,1}}{\delta x_0}\right)^2 + \left(\frac{k_2 - k_1}{\delta k}\right)^2},
\end{equation}
where $\delta x_0$ and $\delta k$ are the grid spacings in $x_0$ and $k$, respectively.  

After GFH identifies candidates, we refine their parameters through a follow-up by making a heterodyne correction which is explained in detail in \cite{2018PhRvD98j2004M}. Using the estimated parameters (initial frequency $f_0$, spin-down $\dot{f}_0$, braking index $n$, and start time $t_0$), we heterodyne the original strain to correct the phase evolution, making the signal nearly monochromatic. This allows longer $T_{\mathrm{FFT}}$ for better frequency resolution. A refined Hough transform then checks signal consistency and updates the CR. Only candidates with improved CR are kept for further analysis, enhancing sensitivity and reducing false alarms.

The GFH algorithm was applied to the post-merger analysis of GW170817 \cite{2019ApJ875160A}, demonstrating its capability to search for signals with significant spin-down, although no detection was made in that search. 

\section{GFH-v2 Methodology}
\label{sec:gfhv2}

\begin{figure*}[ht]
    \centering
    \includegraphics[scale=0.44]{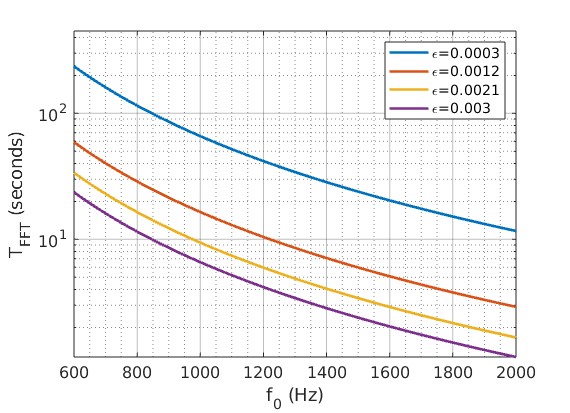}
    \includegraphics[scale=0.43]{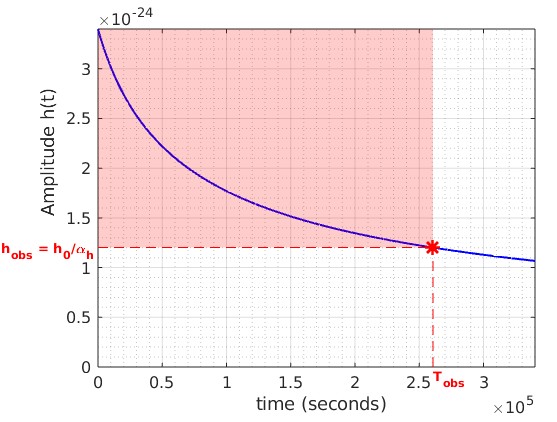}
    \caption{{\it Left:} FFT duration $T_{\rm FFT}$ versus $f_0$ for different ellipticities (see Eq.~\ref{eq:TFFT} - ~\ref{eq:fgw}). {\it Right:} Definition of the observation time $T_{\rm obs}$ from the decay of the signal amplitude for a fixed reduction factor $\alpha_{\rm h}$ shown for $\epsilon = 0.002$ and $f_0 = 900$ Hz.}
    \label{fig:TFFTandTobs}
\end{figure*}
Building upon the original GFH framework, the updated GFH-v2 pipeline introduces a set of methodological and computational improvements in order to enhance both sensitivity and efficiency. The new implementation uses astrophysical motivated parameter constraints together with significant enhancements to the analysis workflow, resulting in a more sensitive, and computationally optimized search framework.

The following subsections summarize the main developments introduced in GFH-v2.

\subsection{Search Parameter Space}
\label{sec:Parameter setup}

\begin{table}
\centering
\caption{Parameter ranges and fixed values used for the injection study.}
\label{tab:parameter_ranges}
\begin{tabular}{l c}
\hline\hline
Parameter & Range/Value \\
\hline
Initial frequency, $f_0$ [Hz] & 600 - 2000 \\
Ellipticity, $\epsilon$ & $3 \times 10^{-4}$ - $3 \times 10^{-3}$ \\
Inclination angle, $\iota$ [deg] & 0 - 180 \\
Braking index, $n$ & 5 \\
Neutron star mass, $M$ [$M_\odot$] & 1.4\\
Neutron star radius, $R$ [km] & 12\\
Sky longitude (RA) [deg] & 210.91 (equatorial) \\
Sky latitude (Dec) [deg] & 54.31 (equatorial) \\
\hline\hline
\end{tabular}
\end{table}

The choice of the search parameter space is guided by both astrophysical considerations and computational feasibility. The properties of a newborn or rapidly rotating neutron star determine the frequency evolution and amplitude of the emitted gravitational waves, while the relative geometry between the source and the detector, such as the sky position of the source and its orientation with respect to the detector, affects the observability at Earth. 

Table~\ref{tab:parameter_ranges} summarizes the key physical parameters relevant to a newborn, rapidly rotating magnetar. The initial frequency \(f_0\), ellipticity \(\epsilon\), and inclination angle \(\iota\) primarily determine the evolution of the signal, including the FFT coherence time \(T_{\mathrm{FFT}}\), the observation time \(T_{\mathrm{obs}}\), and the frequency band spanned by the signal.  
 
The parameter ranges are chosen based on astrophysical expectations. The ellipticity $\epsilon \in [3\times10^{-4},3\times10^{-3}]$ spans plausible values for newly born, rapidly rotating magnetars, motivated by magnetic deformation, residual formation asymmetries, and bar-mode instabilities~\cite{2018MNRAS4801353D,2022ApJ934125X,2021MNRAS5024680S}. The initial frequency $f_0 \in [600,2000]$ Hz covers the early spin-down phase where gravitational-wave emission is strongest~\cite{2020MNRAS4944838L,2021MNRAS5024680S,2023ApJ952156S,2007Ap&SS308119D}. Inclination $\iota \in [0,180^\circ]$ is sampled uniformly in $\cos\iota$, covering all orientations. 
 
 The sky location corresponds to the nearby core-collapse supernova SN 2023ixf, discovered in May 2023 in the Pinwheel Galaxy (M101) at a distance of approximately 6.4 Mpc \cite{Itagaki2023_SN2023ixf,2023ApJ952L23K,2024A&A687L20F}. SN 2023ixf is among the closest Type II supernovae in the past decade, with multi-wavelength observations suggesting the possible formation of a magnetar remnant. Its proximity and well-determined position make it a convenient and astrophysically motivated reference target. The O4a data segment used for the injection study does not overlap with the epoch of the supernova event, ensuring that no real signal from this source is present in the analyzed dataset.

Each parameter combination uniquely determines the expected signal evolution. In particular, the coherence time \(T_{\mathrm{FFT}}\) is set to ensure the signal remains within a single frequency bin (Sec.~\ref{subsubsec:tfft}), while the observation time \(T_{\mathrm{obs}}\) is defined based on the signal amplitude decay (Sec.~\ref{subsubsec:tobs}). The neutron star moment of inertia is computed using Eq.~\eqref{eq:Moment_I}, assuming $M=1.4 M_\odot$ and $R=12$ km.

While Table~\ref{tab:parameter_ranges} defines the representative parameter space for this study, the GFH-v2 methodology is general and can be applied to other astrophysically motivated parameter sets. Details of the signal injection procedure used to validate the pipeline are given in Sec.~\ref{sec:sig_inj}.

\subsubsection{Coherence Time}
\label{subsubsec:tfft}

The coherence time \(T_{\mathrm{FFT}}\) defines the duration of data segments over which we compute Fourier transforms. The choice of \(T_{\mathrm{FFT}}\) depends on the expected signal evolution. It must be short enough that the signal's frequency drift, due to \(\dot{f}_{\rm gw}\), remains within a single frequency bin during that time interval, but long enough to limit sensitivity loss. 
Assuming the frequency drift is approximately linear over an FFT segment, a condition largely verified for the typical segment durations we employ\footnote{In practice, the GFH transform follows the full power-law frequency evolution. The linear approximation using only the first-order spin-down is employed only to set the maximum \(T_{\mathrm{FFT}}\) for each segment. Higher-order terms are negligible over the short segment duration, making this approximation safe.}, the frequency change during \(T_{\mathrm{FFT}}\) is:
\begin{equation}
    \Delta f \simeq |\dot{f}_{\rm gw}| \, T_{\mathrm{FFT}}.
\end{equation}

To keep the signal power within a single frequency bin of width \(1/T_{\mathrm{FFT}}\), we require:
\begin{equation}
    \Delta f \lesssim \frac{1}{T_{\mathrm{FFT}}}.
\end{equation}

Combining the two expressions gives:
\begin{equation}
    |\dot{f}_{\rm gw}| \, T_{\mathrm{FFT}} \lesssim \frac{1}{T_{\mathrm{FFT}}} 
    \quad \Rightarrow \quad
    T_{\mathrm{FFT}} \simeq \frac{1}{\sqrt{|\dot{f}_{\rm gw}|}}.
    \label{eq:TFFT}
\end{equation}
In practice, to ensure that the signal remains confined within one frequency bin throughout the FFT duration, we evaluate $|\dot{f}_{\rm gw}|$ at the start of the signal ($t=t_0$), where the spin-down is largest. This provides a conservative estimate of $T_{\mathrm{FFT}}$.

The spindown rate \(\dot{f}_{\rm gw}\) is related to the ellipticity \(\epsilon\) through Eq.~(\ref{eq:ang_velocity}),
\begin{equation}
    \dot{f}_{gw} = -\frac{32\pi^4 G \epsilon^2 I f_{\rm gw}^5}{5c^5},
    \label{eq:fgw}
\end{equation}
showing that higher ellipticity or frequency results in a faster spin evolution and hence a shorter coherence time. Figure~\ref{fig:TFFTandTobs} left panel illustrates the resulting dependence of \(T_{\mathrm{FFT}}\) on both \(f_0\) and \(\epsilon\).

\subsubsection{Observation time window}
\label{subsubsec:tobs}

In contrast to previous search where the observation time \(T_{\mathrm{obs}}\) is often fixed (see, e.g., \cite{2019ApJ875160A}), we adopt a strategy that optimizes \(T_{\mathrm{obs}}\) based on the signal amplitude decay. Specifically, we define \(T_{\mathrm{obs}}\) as the time it takes for the signal amplitude \(h(t)\) to reduce to a chosen fraction of its initial value \(h_0\), controlled by the amplitude reduction factor \(\alpha_{\mathrm{h}}\):

\begin{equation*}
    T_{\mathrm{obs}} = t \quad \text{s.t.} \quad h(t) = \frac{h_0}{\alpha_{\mathrm{h}}}.
\end{equation*}
Figure~\ref{fig:TFFTandTobs} (right panel) illustrates this definition for a representative case with $f_0 = 900$~Hz, $\epsilon = 0.002$ and a fixed $\alpha_{\rm h}$.  

\begin{figure*}[ht]
    \centering
    \includegraphics[scale=0.44]{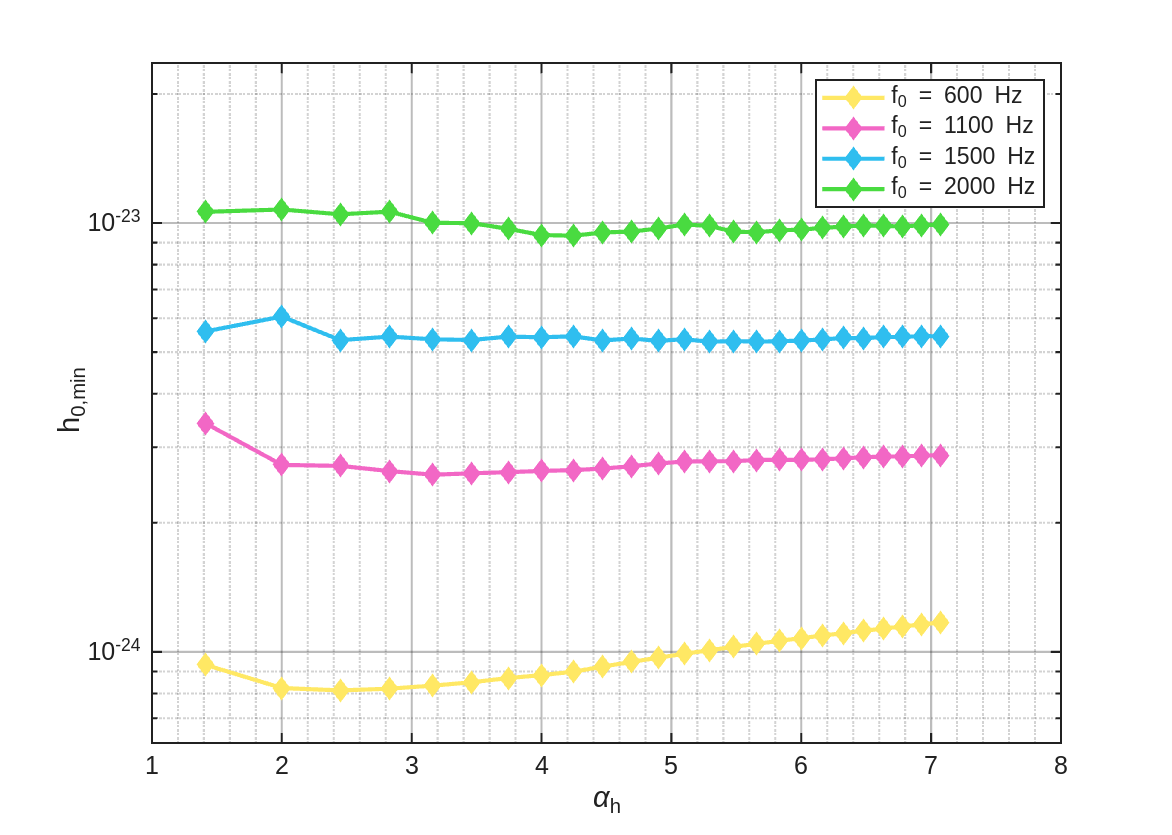}
    \includegraphics[scale=0.44]{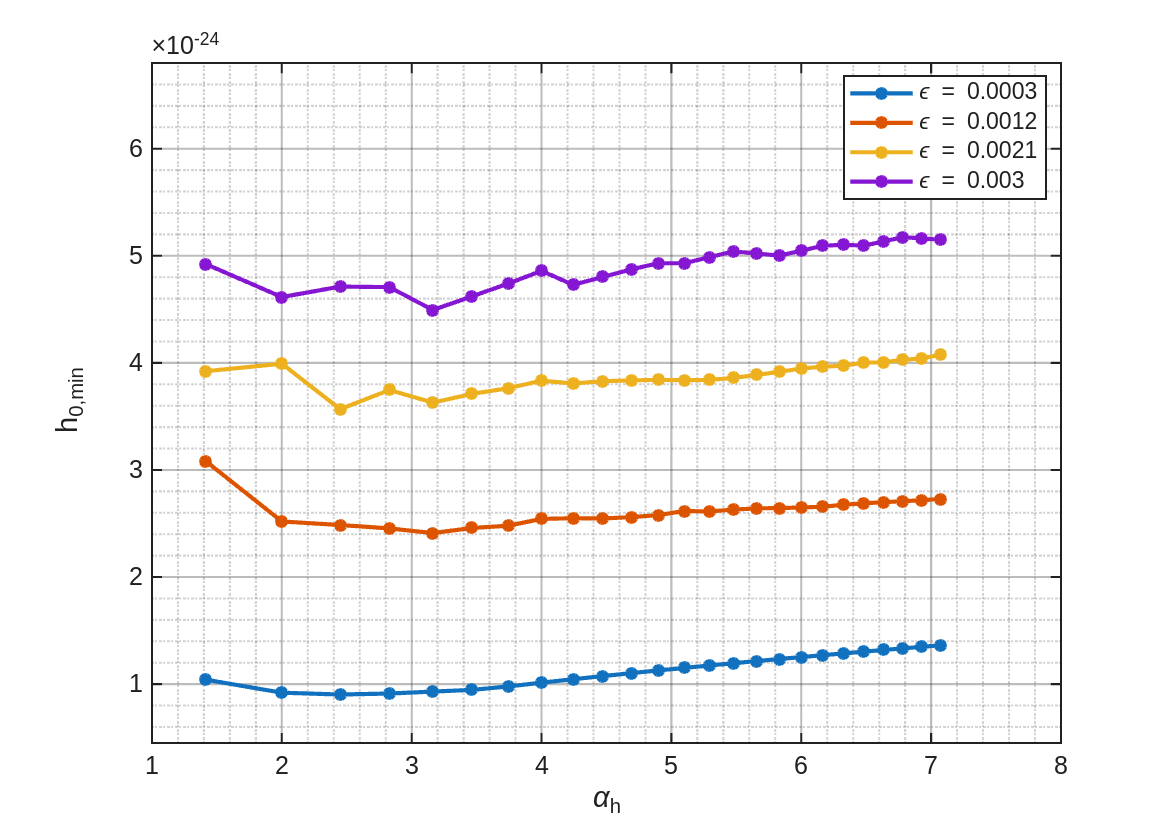}
    \caption{{\it Left:} Sensitivity $h_{\rm 0,min}$ as a function of $\alpha_{\rm h}$ for different initial frequencies $f_0$ and fixed ellipticity $\epsilon = 0.001$.
    {\it Right:} Dependence of $h_{\rm 0,min}$ on $\alpha_{\rm h}$ for different ellipticities $\epsilon$ at a representative $f_0= 1000 Hz$.
    }
    \label{fig:Alpha}
\end{figure*}

To select an optimal value of \(\alpha_{\mathrm{h}}\), we evaluate the sensitivity \(h_{0,\min}\) as a function of \(\alpha_{\mathrm{h}}\) across several representative frequencies and ellipticities, using the formulation described in Section~\ref{sec:theory_sens}(Eq.~\ref{eq:h0min}). As shown in the left panel of Fig.~\ref{fig:Alpha}, the sensitivity initially improves with increasing $\alpha_{\mathrm{h}}$ as more data are integrated, but eventually saturates or worsens as the signal amplitude becomes too weak to contribute effectively. Similarly, the right panel of Fig.~\ref{fig:Alpha} further illustrates the dependence of $h_{\rm 0,min}$ on $\alpha_{\mathrm{h}}$ for different ellipticities $\epsilon$. The optimal $\alpha_{\mathrm{h}}$ is therefore chosen as the value that minimizes $h_{\rm 0,min}$, corresponding to the best achievable sensitivity for a given $f_0$ and $\epsilon$. The variation in $h_{\rm 0,min}$ is moderate and in some cases, shows a weakly defined minimum that determines the optimal value of $\alpha_{\mathrm{h}}$. In cases where the dependence is relatively weak, this result indicates that shorter observation times can be safely used without significantly affecting the sensitivity, thereby reducing computational cost. In other cases the minimum is more pronounced, and selecting the optimal $\alpha_{\mathrm{h}}$ leads to a non-negligible improvement in sensitivity.

\begin{figure}[ht]
    \centering
    \includegraphics[width=0.50\textwidth]{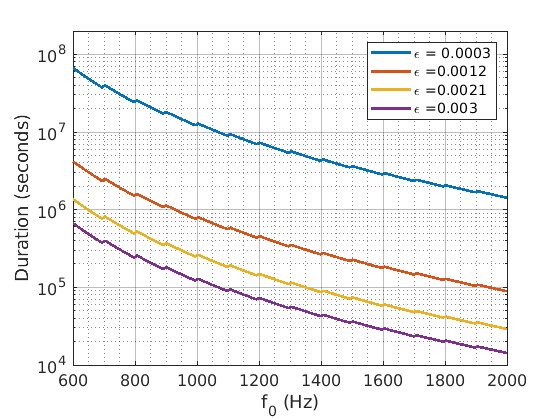}
    \caption{Observation time $T_{\rm obs}$ as a function of $f_0$ for different ellipticities, corresponding to the optimal $\alpha_{\rm h}$ derived from Fig.~\ref{fig:Alpha}.}
    \label{fig:Tobs}
\end{figure}

The observation times $T_{\mathrm{obs}}$ derived from this optimization are shown in Fig.~\ref{fig:Tobs} where each point corresponds to the duration associated with the optimal $\alpha_{\mathrm{h}}$. This approach refines the search strategy by excluding data with negligible signal amplitude, thereby maintaining sensitivity while reducing computational cost.

The frequency band for the analysis is then chosen according to this duration: starting from the initial frequency \(f_0\), we include frequencies down to the value reached at \(T_{\mathrm{obs}}\) due to the spindown corresponding to the chosen \(f_0\) and $\epsilon$. This ensures that only the portion of the data contributing meaningfully to the detection is analyzed. 

\subsection{Implementation of the GFH-v2 algorithm}
\label{subsubsec:GFHv2method}
The GFH-v2 algorithm significantly differentiates from the old GFH~\cite{2018PhRvD98j2004M} and, actually, is completely new for several aspects. One big difference concerns the pre-processing of the detectors' data. Originally, the blocks of FFTs from the SFDB files, which are typically thousands of seconds long, were reprocessed and divided in shorter Fourier-transformed segments with typical length of tens of seconds (see Eq.~\ref{eq:TFFT} for the determination of $T_{\mathrm{FFT}}$) and cut to the desired frequency range. Possible injections of simulated signals for validation purposes were done on that set of reduced FFTs, and the corresponding \textit{peakmap} was created by identifying significant local maxima in the normalized power spectrum of each FFT segment (see Sec.~\ref{sec:gfh}). In the GFH-v2, the blocks from the SFDB are directly combined together and inverse Fourier transformed to create a sub-sampled, complex reduced-analytic time series that covers the frequency range spanned by the signal during $T_{\mathrm{obs}}$ ( from \(f_0\) down to the value reached due to spin-down, see Sec.~\ref{subsubsec:tobs}). This format is analogous to the Band-sampled data (BSD) \cite{BSD_Piccinni_2019}, widely used in CW searches. However, the additional cleaning steps applied in standard BSD, which are designed to remove persistent narrow-band artifacts in quasi-monochromatic CW searches, are not applied here, since the signal evolves over a broader frequency range. This approach allows all the manipulations, like injection of signals or generation of the \textit{peakmap}, to be done easily exploiting the existing libraries and making the code easily accessible to the gravitational-wave community~\cite{Pierini2025GFHv2}.

The other big change concerns the implementation of the core of the GFH-v2 transform itself. In the original GFH, each peak from the \textit{peakmap} was transformed into a line in the Hough space through an algorithm based on a double nested loop. The first loop spans between the different times in the \textit{peakmap}, selecting the peaks happening at those times, while the second loop writes on the Hough map spanning through all the values of the $k$ parameter according to a discrete grid. The main limitations of this algorithm are its computational efficiency and the fact the loops for different peaks could access the same memory locations for writing, limiting the possibility to parallelize the operation and to exploit modern computing architectures. In the GFH-v2 implementation, the loop over time is removed. The loop now runs over the grid in $k$: each value corresponds to a different slope of the signal in the linearized \textit{peakmap}, according to Eq. \ref{eq:gfh_line}. For each $k$ value, the whole \textit{peakmap} is integrated along the corresponding slope by combining one linear transformation and the computation of an histogram along the grid of transformed frequencies $x_0$. In this study, the histogram counts are unweighted (each peak contributes a value of one); the use of weights is planned for future improvements.

The result of the histogram fills one column in the Hough map at the location corresponding to the selected $k$ value. This new implementation of the transform is made by means of highly-optimized functions, which compute the 1D histograms using buffer arrays to speed up memory access. Compared to the original implementation, where the Hough map was computed through a double nested loop over time indexes and $k$ values with direct memory access, the GFH-v2 avoids the outer time loop and efficiently integrates the entire peakmap along each $k$ value. Quantitative tests show that the computing time is reduced by approximately one order of magnitude. A detailed comparison of the old and new implementations, including computing times as a function of the number of cores, is provided in Appendix~\ref{app:computing_speed}. The implementation details and the actual code for the new GFH-v2 are documented in~\cite{Pierini2025GFHv2}. Moreover, each different $k$ value of the grid is mapped in a different column in the Hough map, so the loop can be performed in parallel exploiting hyper-threading or multi-core CPUs. While the pipeline can also be executed on Graphical Processing Units (GPUs)~\cite{Rosa:2021ptb}, the current search is efficiently performed on standard CPU resources; GPU execution is available if needed for future searches with higher computational demands.

\begin{figure}[ht]
    \centering
    \includegraphics[width=0.3\textwidth]{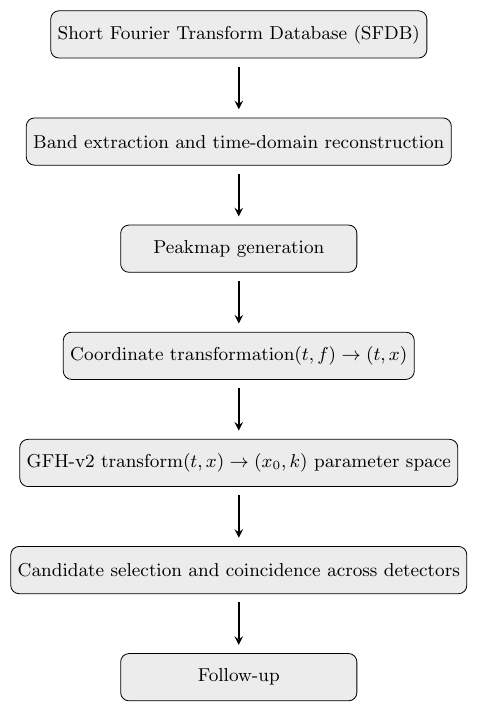}
    \caption{Schematic view of the GFH-v2 pipeline workflow.}
    \label{fig:workflow}
\end{figure}

A schematic representation of the complete GFH-v2 workflow is shown in Fig.~\ref{fig:workflow}, summarizing the main processing steps from the input SFDB to the follow-up~\cite{Pierini2025GFHv2} (see Sec.~\ref{sec:gfh} and Ref.~\cite{2018PhRvD98j2004M} for details on the heterodyne-based candidate refinement).

\begin{figure*}[ht]
    \centering
    \includegraphics[scale=0.45]{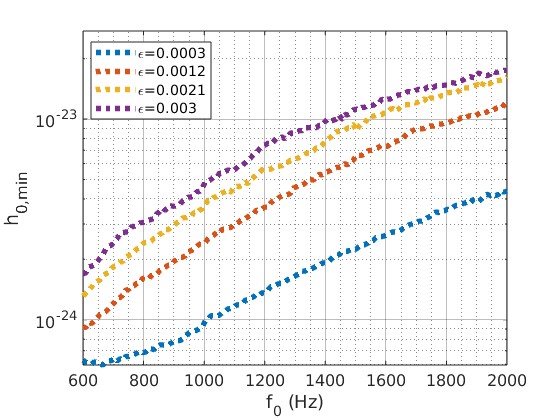}
    \includegraphics[scale=0.45]{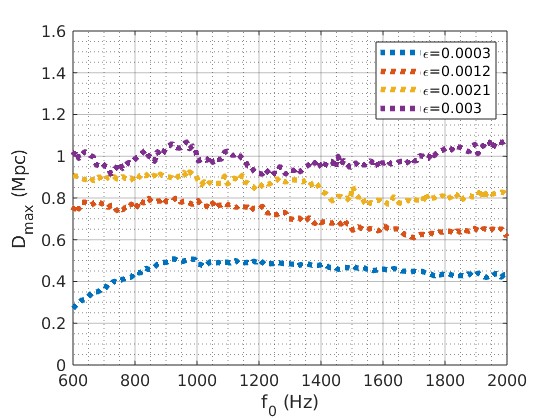}
    \caption{Theoretical sensitivity obtained using full O4a LIGO data for the Livingston and Hanford detectors~\cite{2025arXiv250818079T,GWOSC}, taking the maximum PSD value between the two detectors at each frequency to ensure a conservative estimate (see also Fig.~\ref{fig:O4a} (left panel) for the PSD overview)
    {\it Left:} Minimum detectable strain amplitude $h_{\mathrm{0,min}}$ as a function of the signal initial frequency $f_0$. 
    {\it Right:} Corresponding maximum detectable distance $D_{\mathrm{max}}$ as a function of $f_0$. Both panels refer to $\Gamma = 90$ and a critical ratio threshold of $CR_{\mathrm{thr}} = 7$.}
    \label{fig:TheoreticalSensitivity}
\end{figure*}

\section{Theoretical Sensitivity}
\label{sec:theory_sens}
Theoretical sensitivity provides a baseline to evaluate the performance of the GFH-v2 pipeline and to quantify the impact of detector noise and analysis parameters. For tCW searches from newborn magnetars, the standard CW sensitivity expression can be generalized to account for the rapidly evolving frequency. The resulting minimum detectable strain is derived in \cite{Peaksens}:  

\begin{equation}
       h_{0,\mathrm{min}} = \mathcal{A}_{\rm min} f_0^2 \\
       \label{eq:h0min}
\end{equation}
where, 
\begin{equation}
    \begin{split}
        \mathcal{A}_{\rm min} =\sqrt{\frac{2 \pi}{2.4308}} \frac{N_{\mathrm{eff}}^{1/4}}{\sqrt{T_{\mathrm{FFT}}}} \left(\frac{p_{0} (1-p_{0})}{p_{1}^{2}}\right)^{1/4}\\
        \times \left(\sum_{i=1}^{N_{\rm seg}} \frac{f_i^4 ( F_+ A_+ + F_\times A_\times)^2_i}{S_n(f_i)} \right)^{-1/2} \\
        \times \sqrt{CR_{\mathrm{thr}} - \sqrt{2}\, \mathrm{erfc}^{-1} (2\Gamma)},
    \end{split}
    \label{eq:Amin}
\end{equation}

Note that $\mathcal{A}_{\rm min}$ carries units of $\rm Hz^{-2}$, which compensate the explicit $f_0^2$ factor in Eq.~\eqref{eq:h0min}, ensuring that the strain amplitude $h_{0,\mathrm{min}}$ remains dimensionless. Here, $N_{\mathrm{eff}} = T_{\mathrm{obs,eff}} / T_{\mathrm{FFT}}$ is the number of effective segments, where $T_{\mathrm{obs,eff}}$ is the total duration of science data within the selected observation time $T_{\mathrm{obs}}$~\cite{GWOSC}. With interlaced FFTs, the total number of FFTs is $N_{\rm seg} = 2 N_{\mathrm{eff}}$. The quantity $p_0$ represent the probabilities of selecting a noise peak above a peak-selection threshold $\theta$ (typically $\theta = 2.5$~\cite{2005CQGra22S1255P}) applied to the normalized periodogram in each FFT, whereas $p_1$ represents the linear change in the peak-selection probability when a weak signal of amplitude $\lambda$ is present ( $p_\lambda = p_0 + p_1 \lambda$). They are given by
\begin{equation}
p_0 = e^{-\theta} - e^{-2\theta} + \frac{1}{3}e^{-3\theta}
\label{eq:p0}
\end{equation}
and  
\begin{equation}
p_1 = \theta \bigl(\frac{1}{2} e^{-\theta} - \frac{1}{2} e^{-2\theta} + \frac{1}{6}e^{-3\theta} \bigr) + \frac{1}{4} e^{-2\theta} - \frac{1}{9} e^{-3\theta}
\label{eq:p1}
\end{equation}

The sum in Eq.~\eqref{eq:Amin} extends over all $N_{\rm seg}$ time segments where $f_i$ denotes the instantaneous signal frequency\footnote{In previous works~\cite{Peaksens, 2018PhRvD98j2004M}, the quantity $\mathcal{F}_i$ was introduced to denote the scaling of the signal amplitude with frequency. For newborn magnetars, the signal amplitude scales as $f_i^2$, so we directly replace $\mathcal{F}_i^2$ with $f_i^4$ here for clarity.}. The factors $F_+$ and $F_\times$ are the detector response functions~\cite{1998PhRvD58f3001J}, while $A_+$ and $A_\times$ are defined as $A_+ = \frac{1 + cos^2(\iota)}{2}$ and $A_\times = cos(\iota)$. The noise power spectral density is denoted by $S_n$. 

The factor $CR_{\mathrm{thr}}$ defines the critical ratio threshold for candidate selection, balancing sensitivity and false-alarm probability. In our tests with simulated signals, we adopt $CR_{\mathrm{thr}} = 7$, which provides a good compromise between the two (see Appendix~\ref{app:crthr}). While in principle the optimal $CR_{\rm thr}$ could depend on analysis parameters such as $T_{\rm obs}$ and $T_{\rm FFT}$ due to differences in the number of effective trials, a simple theoretical estimate is difficult. We therefore determined the threshold empirically via injection studies, ensuring a false-alarm probability below 1\% across the parameter space. In a real search, $CR_{\rm thr}$ could be adjusted for different follow-up strategies, but always following the same procedure to control the false-alarm rate. Finally, $\Gamma$ denotes the chosen confidence level, i.e., the fraction of repeated trials in which a signal with amplitude $h_{0,\mathrm{min}}$ would be recovered, and $\mathrm{erfc}^{-1}$ is the inverse complementary error function.

The corresponding maximum distance can be is obtained by equating the gravitational-wave strain, Eq. \ref{eq:amplitude}, to the sensitivity, Eq. \ref{eq:h0min}, obtaining
\begin{equation}
D_{\rm max} = \frac{4 \pi^2 G}{c^4} \frac{I \, \epsilon}{\mathcal{A}_{\rm min}},
\label{eq:dmax}
\end{equation}

\begin{figure*}[ht]
    \centering
    \includegraphics[width=0.49\textwidth]{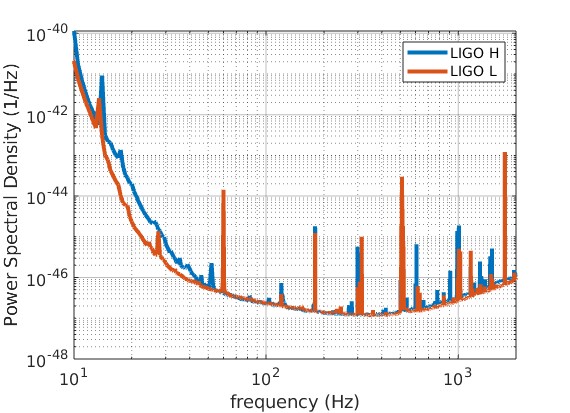}
    \includegraphics[width=0.49\textwidth]{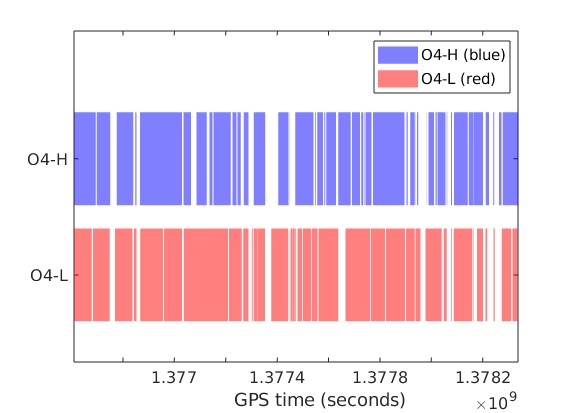}
    \caption{Overview of the LIGO O4a dataset used in this study. 
    {\it Left:} Average noise spectral densities of both detectors computed over the full O4a dataset. {\it Right:} Science segments from the two LIGO detectors (Hanford and Livingston) over the 20-day analysis window starting on 20 August 2023 which is used for the injection studies in Sec.\ref{sec: EFfandsens}.}
    \label{fig:O4a}
\end{figure*}
Figure~\ref{fig:TheoreticalSensitivity} shows the theoretical sensitivity of the GFH-v2 pipeline computed using these equations, with real O4a data~\cite{GWOSC,2025arXiv250818079T} and actual search parameters as inputs. The left panel shows the minimum detectable strain $h_{0,\mathrm{min}}$, while the right panel shows the corresponding maximum distance $D_{\rm max}$ as a function of the initial frequency $f_0$ for a few representative ellipticities. For this estimate we use power spectral densities (PSD) derived from the full O4a dataset for both LIGO detectors. The PSDs are computed from the SFDB data~\cite{Astone2025SFDB} by averaging the power spectra of the individual FFT segments belonging to science time, applying basic data-quality vetoes and a weighted averaging procedure that reduces the influence of noisier spectra~\cite{2014PhRvD90d2002A}. At each frequency we then take the maximum PSD value between LIGO Hanford and LIGO Livingston to obtain a conservative sensitivity estimate~\footnote{This calculation provides a general reference sensitivity of the pipeline using the full O4a dataset and is not tied to the specific data segment used in the injection studies presented in Sec.~\ref{sec:exp_sens}.}

The trend of these plots can be understood from the underlying sensitivity equations Eq.~\eqref{eq:h0min}-\eqref{eq:dmax}. The curve for $h_{0,\rm min}$ generally increases with frequency due to the $f_0^2$ scaling in Eq.~\eqref{eq:h0min}. However, the overall trend results from a combination of competing effects: higher ellipticity increases the signal amplitude but also shortens the optimal coherence time $T_{\mathrm{FFT}}$ (see Sec.~\ref{subsubsec:tfft}), which degrades the sensitivity. The net outcome of these opposing factors is the behavior observed in the plots. The impact of the real detector PSDs, which are not flat, introduces small frequency-dependent variations in the sensitivity, producing small-scale fluctuations in the curves. These PSD-induced variations are smaller than the main $f_0^2$-driven trend and do not qualitatively alter the observed sensitivity behavior.

\section{Empirical Sensitivity}
\label{sec:exp_sens}
The theoretical sensitivity derived in the previous section assumes data follow a Gaussian distribution. Although this assumption can be reasonable to a first approximation, typical detector data often show non-Gaussianity both in time and in frequency. It is thus important to estimate the search sensitivity in a more empirical way, which takes in full consideration the characteristics of the data.  

To empirically estimate the sensitivity of the pipeline, we performed simulated signal injections into O4a data across representative values of frequency and ellipticity. For each injection set, we determined the minimum detectable strain amplitude (or equivalently, the maximum distance) as that strain such that 90$\%$ of the signals were successfully recovered.

\subsection{O4a Data Characteristics}
\label{sec:O4adata}

Our test dataset is drawn from the first portion of the LIGO-Virgo-KAGRA fourth observing run (O4), hereafter O4a, spanning 24 May 2023 to 16 January 2024~\cite{2025arXiv250818079T,GWOSC}.  
Relative to O3, O4a marked a significant advance in instrument performance, with substantial upgrades to the interferometers~\cite{2025PhRvD.111f2002C}. 

During this period, the Virgo detector~\cite{2015CQGra32b4001A} was not operating due to commissioning activities aimed at improving its sensitivity. KAGRA~\cite{2021PTEP2021eA102A} and GEO 600~\cite{2002CQGra191377W} were operating but remained less sensitive than the LIGO instruments~\cite{2015CQGra32g4001L} and thus, the present study uses only data from the two LIGO detectors (Hanford and Livingston). Figure~\ref{fig:O4a} left panel shows the average noise power spectral densities of the two detectors computed over the full O4a dataset.

We selected a 20-day window starting from Aug 20, 2023 (MJD: 60176.991) as the input data for the empirical sensitivity evaluation. The GFH-v2 pipeline determines the effective observation time $T_{\mathrm{obs}}$ individually for each signal, based on its initial frequency $f_0$ and ellipticity $\epsilon$ (see Fig.~\ref{fig:Tobs}). Only this effective duration is used in the analysis. For the upper ellipticity sub-interval considered in the empirical sensitivity study (Sec.~\ref{sec: EFfandsens}), the chosen 20-day segment fully contains the evolution of the signals. The science segments within this window~\cite{GWOSC} are shown in the right panel of Fig.~\ref{fig:O4a}.

\subsection{Signal Injections}
\label{sec:sig_inj}
We performed simulated signal injections into the time-domain reduced-analytic series derived from O4a SFDB data, as described in Sec.~\ref{subsubsec:GFHv2method}, using the parameter ranges defined in Sec.~\ref{sec:Parameter setup} (Table~\ref{tab:parameter_ranges}). The full ellipticity range ($\epsilon \in [3\times10^{-4},3\times10^{-3}]$) was divided into 5 representative sub-intervals, each spanning comparable logarithmic intervals to follow the gradual change in signal amplitude and spin-down behaviour across the full range while keeping the computational cost manageable. For each sub-interval, and at a fixed source distance, injections were carried out in steps of 10~Hz, with approximately 150 signals uniformly distributed in each frequency interval across the full $f_0$ range\footnote{The exact number of injections varies between 120 and 180, depending on the frequency band; additional injections were performed in certain frequency intervals.}. The ellipticity for each signal is drawn randomly from a uniform distribution within the selected sub-interval. Other physical parameters, including the inclination angle and polarization angle, are sampled from uniform distributions within their respective ranges (e.g., \(\cos\iota\) uniformly in [-1,1]). This procedure was repeated for different source distances to trace the detection horizon. 

For the empirical sensitivity evaluation (Sec.~\ref{sec: EFfandsens}), only the upper ellipticity sub-interval ($\epsilon \in [2.5\times10^{-3},3\times10^{-3}]$) is used, corresponding to the strongest signals. The effective observation time $T_{\mathrm{obs}}$ for each injected signal is determined by its $f_0$ and $\epsilon$ (see Fig.~\ref{fig:Tobs}), ensuring that the signal is fully contained within the 20-day O4a data window described in Sec.~\ref{sec:O4adata}. The FFT coherence times $T_{\mathrm{FFT}}$ were set consistently with the definitions in Sec.~\ref{sec:Parameter setup}. 

All injections are assumed to start at the beginning of the dataset for simplicity. In more realistic scenarios, the signal start time may not coincide with the beginning of the analyzed data. Allowing for different starting times introduces an additional trials factor, which can slightly reduce sensitivity. In future work, we plan to scan over a discrete grid of possible start times, choosing a step size that balances parameter-space coverage, computational cost, and minimal sensitivity degradation.

\subsection{Candidate Selection}

The injected datasets were analyzed with GFH-v2, and candidates were selected from the resulting Hough maps. To systematically cover the parameter space, the Hough map was divided as follows: the $x_0$ parameter was divided into a number of blocks determined by the number of injections and width of the spindown band covered by the signal\footnote{In the injection study, the number of blocks along the $x_0$ axis was scaled with the number of injected signals to control the candidate density. In a real search, a fixed partitioning scheme is used instead (e.g., 100 blocks in $x_0$).}. The $k$ axis was divided such that each block is approximately square-shaped, containing the same number of bins along $x_0$ and $k$. Within each block the bin with the highest map count was selected as a candidate. A second candidate was chosen only if it was sufficiently far away from the first (in this case, by 6 bins in $x_{0}$) to reasonably select candidates of different origin.

Coincidence checks were performed across detectors and with respect to the injected parameters to identify common candidates. Since only $x_0$ and $k$ were varied in this study, coincidences were evaluated using a threshold of $d_{\mathrm{coin}} = 2$. In previous implementations, additional parameters such as braking index and sky position were included in the search, requiring a higher threshold.

\subsection{Detection Efficiency and Sensitivity}
\label{sec: EFfandsens}
Following the candidate selection, we determine the detection efficiency for each signal initial frequency, defined as the fraction of injected signals that are successfully recovered with $CR \ge CR_{\mathrm{thr}}$ and coincident across the detectors (see Appendix~\ref{app:crthr} for details on how $CR_{\rm thr}$ was chosen using injection tests on real O4a data). 

For this analysis, we restrict the study to the upper ellipticity sub-interval ($\epsilon \in [2.5\times10^{-3},3\times10^{-3}]$) defined in Sec. \ref{sec:sig_inj}, which corresponds to the strongest signals in the considered parameter space. Lower ellipticities result in proportionally smaller detectable distances without significantly affecting the efficiency curves. We note that averaging over this small ellipticity sub-interval produces only minor differences in the resulting sensitivity estimates\footnote{In practice, results are always evaluated separately for each ellipticity sub-interval rather than averaged over the full range ($\epsilon \in [3\times10^{-4},3\times10^{-3}]$), ensuring that the dependence on $\epsilon$ is properly captured. Here, we present the sensitivity for the upper sub-interval to illustrate the maximum detection horizon achievable by the pipeline. In a real search, upper limits are also computed separately for each ellipticity sub-interval.}.

To estimate the empirical maximum detectable distance, $D_{\mathrm{max}}$, we identify the point at which the detection efficiency curve crosses the 90\% recovery level. For each frequency, injections are sorted by increasing distance and their corresponding efficiencies are evaluated. The first distance value at which the efficiency falls below 90\% is taken as the upper bracketing point, while the immediately preceding distance value, that yields an efficiency above 90\%, is taken as the lower bracketing point. The crossing point, $D_{\mathrm{max}}$, is then obtained by linear interpolation between these two values. This procedure is based directly on the discrete set of injection results and does not assume any particular functional form for the efficiency curve. Although one could estimate the 90\% crossing using a smooth fit to the data, the approach adopted here avoids extrapolation beyond the data points and provides a simple and data-driven estimate. The associated uncertainty on $D_{\mathrm{max}}$ is derived from the spacing between these neighboring distance points, defining asymmetric error bars.
\begin{figure}[ht]
    \centering
    \includegraphics[width=0.50\textwidth]{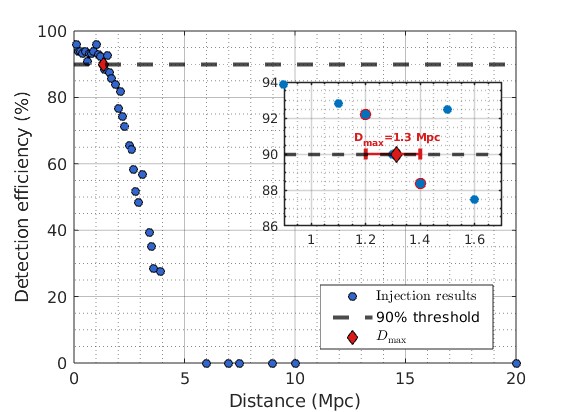}
    \caption{Detection efficiency curve for a representative frequency ($f_0 = 1400$~Hz) in the upper ellipticity sub-interval, $\epsilon \in [2.5\times10^{-3},\,3\times10^{-3}]$, using $CR_{\mathrm{thr}} = 7$. 
    The maximum detectable distance, $D_{\mathrm{max}}$, is determined by linear interpolation at the 90\% recovery level, with horizontal bars representing the errors derived from the bracketing distances.}
    \label{fig:efficiency_curve}
\end{figure}

Figure~\ref{fig:efficiency_curve} illustrates a representative example of a detection efficiency curve for a single frequency, showing how $D_{\mathrm{max}}$ and its associated error.

\begin{figure}[ht]
    \centering
    \includegraphics[width=0.5\textwidth]{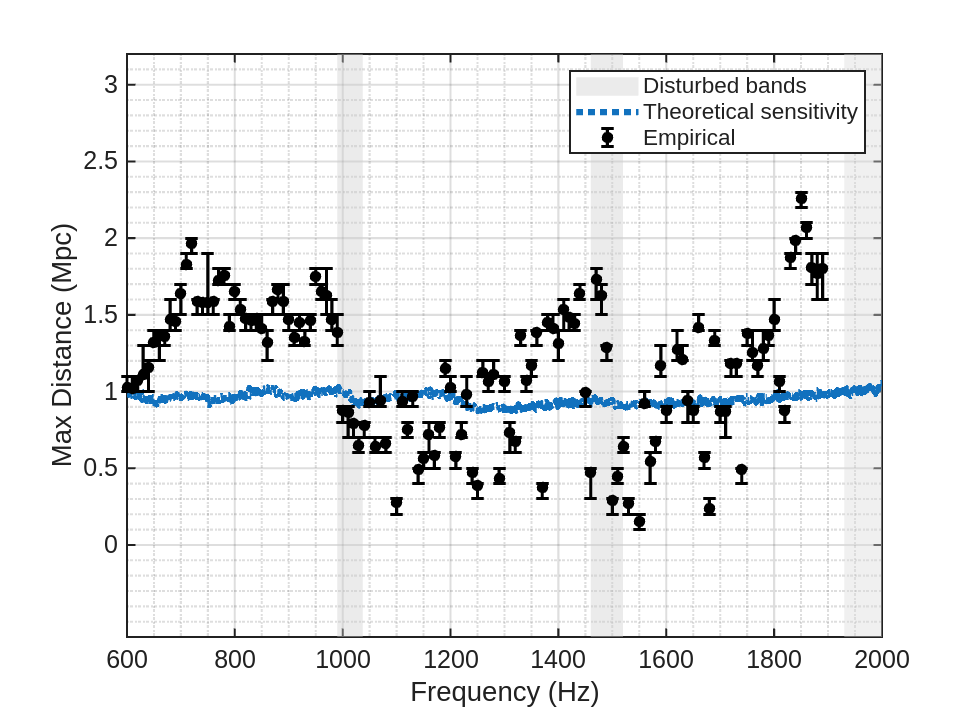}
    \caption{
    Comparison between the empirical maximum distance obtained from signal injections and the theoretical sensitivity estimate across the analyzed frequency range, for the upper ellipticity sub-interval $\epsilon \in [2.5\times10^{-3},\,3\times10^{-3}]$, using a 90\% confidence level and $CR_{\mathrm{thr}} = 7$. The theoretical sensitivity here uses a 10-day PSD corresponding to the effective duration of the injections in the upper ellipticity sub-interval, taking the maximum value between H1 and L1 detectors to remain conservative. The shaded gray regions indicate frequency bands affected by instrumental disturbances~\cite{GWOSC_O4_SpecLines}, which lead to a degradation of the empirical sensitivity.}
    \label{fig:sensitivity_comparison}
\end{figure}

Figure~\ref{fig:sensitivity_comparison} presents the comparison between the empirical maximum detectable distances obtained from injection studies and the corresponding theoretical sensitivity estimates across the analyzed frequency range. Overall, the empirical sensitivity follows the theoretical trend, although deviations up to a factor of $\sim 2$ are observed in certain frequency bands. In some bands, the empirical $D_{\mathrm{max}}$ falls below the theoretical prediction, typically in regions affected by strong spectral features, such as suspension violin modes and their harmonics~\cite{GWOSC_O4_SpecLines} (as highlighted by the shaded gray regions in Fig.~\ref{fig:sensitivity_comparison}). These disturbances raise the local noise level and reduce the recovery efficiency, leading to a lower measured sensitivity relative to the theoretical expectation.

Conversely, at some frequencies, the empirical sensitivity exceeds the theoretical one, sometimes by factors of 1.5-2. One possible explanation is that the theoretical sensitivity curve is based on a conservative assumption: at each frequency, it uses the maximum of the H1 and L1 detector PSDs. This choice effectively corresponds to adopting the noisier of the two detectors at that frequency, and therefore provides a conservative estimate of the noise level. As a result, the predicted sensitivity is slightly underestimated.

Overall, the comparison indicates that the empirical results are consistent with theoretical expectations at the order-of-magnitude level, capturing the main trends while naturally reflecting real data variations.

\section{Conclusion}
\label{sec:concl}
In this work, we have presented an enhanced methodology for the detection of tCW signals using the updated GFH-v2 pipeline. This approach introduces astrophysically motivated parameter constraints, optimized choices of coherence and observation times, and computational improvements that significantly enhance both sensitivity and computational efficiency. Using real O4a LIGO data and simulated tCW signals describing newborn magnetar emission, we demonstrated the pipeline’s ability to recover signals over a broad parameter space and to produce empirical sensitivities consistent with theoretical expectations at the order-of-magnitude level. These results establish GFH-v2 as a robust and efficient framework for directed searches for tCW signals in current and future observing runs.

The pipeline has already been applied in a real directed search targeting the nearby core-collapse supernova SN2023ixf, detected in May 2023. This analysis was carried out using data from LIGO Engineering Run 15 (ER15), which was conducted shortly before the start of the O4 observing run. We used the subset of ER15 data segments approved for LVK publications~\cite{GWOSC}. The source localization provided by electromagnetic observations enables a well-constrained directed search using the framework developed in this study. Results from that analysis will be presented in a forthcoming publication.

In future work, we plan to extend this study in several directions. First, we will relax the assumption that all signals start at the same time and test different start times to see how this affects detection efficiency. We also plan to include a wider range of magnetar spin-down models, considering different braking indices \cite{2015PhRvD91f3007H} and possible fallback accretion effects, to check how robust the method is under various physical conditions \cite{2009MNRAS3981869D,2012ApJ76163P,2011PhRvD83j4014C}. Additionally, we plan to implement an adaptive Hough approach, where the contribution of each time segment is weighted according to the detector noise level and the beam pattern, following the method described in Ref.~\cite{2014PhRvD90d2002A,2019PhRvD99j4067O}. 

Moreover, a promising direction is to develop a hybrid pipeline that combines GFH-v2 with a machine-learning (ML) based tCW search. Recent studies have demonstrated that convolutional neural networks can efficiently identify tCW signals and generate reliable triggers at very low computational cost ~\cite{2024PhRvD110j3047A,2019PhRvD.100f2005M}. In this framework, the ML pipeline can rapidly scan large portions of parameter space to generate candidate triggers, which are then followed up by GFH-v2 for a detailed and sensitive analysis. This approach could enable efficient all-sky searches for long transient signals, combining speed and sensitivity.

Finally, the GFH-v2 framework can also be applied to search for other long-duration sources, such as post-merger remnants, using data from future observing runs with the advanced LIGO, Virgo, and KAGRA detectors.

Overall, the results presented here demonstrate that the GFH-v2 pipeline provides a sensitive, efficient, and astrophysically informed framework for long-transient gravitational-wave searches, representing an important step toward probing the physics of newly born magnetars and other rapidly evolving compact objects with gravitational waves.

\begin{acknowledgments}
This work is partially supported by ICSC - Centro Nazionale di Ricerca in High Performance Computing, Big Data and Quantum Computing, funded by European Union - NextGenerationEU.

This research has made use of data or software obtained from the Gravitational Wave Open Science Center (gw-openscience.org), a service of LIGO Laboratory, the LIGO Scientific Collaboration, the Virgo Collaboration, and KAGRA. LIGO Laboratory and Advanced LIGO are funded by the United States National Science Foundation (NSF) as well as the Science and Technology Facilities Council (STFC) of the United Kingdom, the Max-Planck-Society (MPS), and the State of Niedersachsen/Germany for support of the construction of Advanced LIGO and construction and operation of the GEO600 detector. Additional support for Advanced LIGO was provided by the Australian Research Council. 

Virgo is funded, through the European Gravitational Observatory (EGO), by the French Centre National de Recherche Scientifique (CNRS), the Italian Istituto Nazionale di Fisica Nucleare (INFN) and the Dutch Nikhef, with contributions by institutions from Belgium, Germany, Greece, Hungary, Ireland, Japan, Monaco, Poland, Portugal, Spain. 

The construction and operation of KAGRA are funded by Ministry of Education, Culture, Sports, Science and Technology (MEXT), and Japan Society for the Promotion of Science (JSPS), National Research Foundation (NRF) and Ministry of Science and ICT (MSIT) in Korea, Academia Sinica (AS) and the Ministry of Science and Technology (MoST) in Taiwan.

We also thank the Amaldi Research Center, for the clusters hosted in Rome INFN, where we have stored the LIGO/Virgo data used in this research and run part of the present analysis. We thank the INFN-CNAF computing staff for the resources we have used in this analysis and for their constant support. 

We further acknowledge Iuri La Rosa and Luca Rei for their constructive review and insightful feedback for the pipeline.

\end{acknowledgments}
\section*{Data Availability}
The data that support the findings of this article are
publicly available~\cite{GWOSC,2025arXiv250818079T}. 

\appendix

\section{Quantitative Comparison of GFH Implementations}
\label{app:computing_speed}

\begin{figure}[ht]
    \centering
    \includegraphics[width=0.53\textwidth]{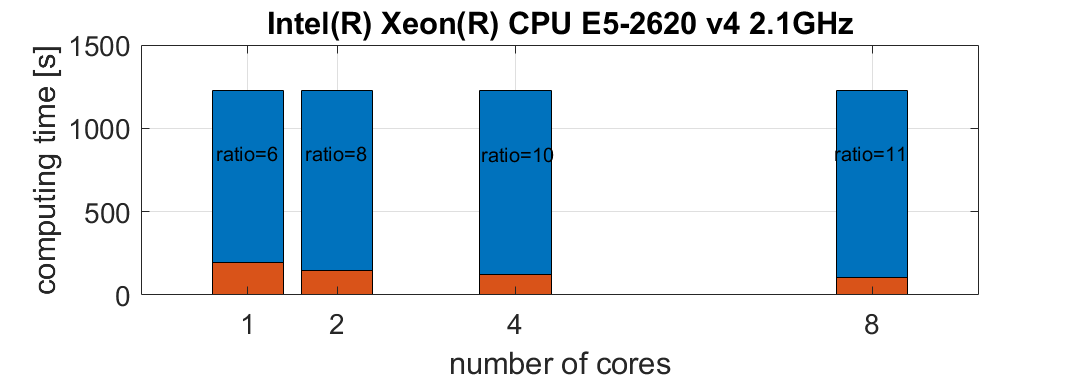}
    \caption{Computing time comparison between the old GFH and GFH-v2 implementations. Blue bars: old code (single-thread). Red bars: GFH-v2 (multi-threaded).}
    \label{fig:GFH_speed}
\end{figure}

Figure~\ref{fig:GFH_speed} shows a comparison of the computing times for the original GFH and the GFH-v2 implementations. The test dataset covers an 800 Hz-wide frequency band and a 9-hour time window, analyzed with $T_{\rm FFT}=2$~s. The tests were performed on an Intel Xeon E5-2620 machine (16 cores, 2.1 GHz, hyperthreading). While the original implementation runs on a single thread, GFH-v2 exploits multithreading. The blue and red bars indicate the computing time for the old and new codes, respectively, as a function of the number of cores used. The results demonstrate that GFH-v2 achieves a speedup of roughly one order of magnitude compared to the original implementation.

\section{Theoretical sensitivity equation}
\label{app:theosens}
In this appendix, we give some details on the derivation of the sensitivity formula, Eq. \ref{eq:Amin}, directing the reader to \cite{Peaksens} for more details.
The sensitivity computation consists of several steps and starts from two basic quantities: the probability of selecting a noise peak (local maxima above the threshold $\theta$) in the peakmap, $p_0=P(\theta;0)$, given by Eq. \ref{eq:p0}, and the probability of selecting a peak when a signal with spectral amplitude $\lambda$ (in units of the equalized spectra) is present, $p_\lambda=P(\theta;\lambda)$. Details of the original computation, for CW signals, can be found in \cite{2014PhRvD90d2002A}, while its extension to long-transient signals is given in \cite{2018PhRvD98j2004M}. Under the assumption of Gaussian noise, the probability of selecting a peak in the peakmap in presence of a signal with spectral amplitude $\lambda$ is
\begin{equation}
    p_\lambda = \int_\theta^\infty p(x;\lambda) \left(\int_0^x e^{-x'}dx'\right)^2dx
    \label{eq:plambda}
\end{equation}
where $x$ is the value of the noise spectrum.
This equation corrects an error in \cite{2014PhRvD90d2002A} (eq. 19), where the inner integral had as argument $p(x;\lambda)$, instead of $p(x;0)$. By computing the integral of Eq. \ref{eq:plambda}, under the small signal approximation, we find
\begin{equation}
    p_\lambda = p_0 + p_1\lambda
\end{equation}
where $p_0$ is given by Eq. \ref{eq:p0} and 
\begin{equation}
p_1 = \theta \left(\frac{1}{2}e^{-\theta} - \frac{1}{2}e^{-2\theta} + \frac{1}{6}e^{-3\theta}\right) + \frac{1}{4}e^{-2\theta} - \frac{1}{9}e^{-3\theta},
\end{equation}
which corresponds to Eq. \ref{eq:p1}.
The expression for the minimum detectable $\lambda$ now reads (cfg. eq 66 in \cite{2014PhRvD90d2002A})
\begin{equation}
\lambda_\mathrm{min}=\sqrt{\frac{p_0(1-p_0)}{N_\mathrm{eff}\times p_1^2}}\left(CR_\mathrm{thr} -\sqrt{2}\mathrm{erfc}^{-1}(2\Gamma)\right).
\end{equation}

By following the same steps as in \cite{2014PhRvD90d2002A}, and considering the case of a signal with finite duration, we arrive at the expression for the theoretical sensitivity of the GFH-v2 algorithm, given in Eq. \ref{eq:Amin}. Note that Eq. \ref{eq:Amin} replaces eq. 34 of \cite{2018PhRvD98j2004M}, not only because of the difference in the expression for the coefficient $p_1$ but also because in Eq. \ref{eq:Amin} the dependence on the detector radiation pattern is explicit, while in \cite{2018PhRvD98j2004M} an average over time (and the source parameters) is assumed. This is not correct when the signal duration is smaller than about one sidereal day, in which case no time average can be taken.     

\section{Selection of the Critical Ratio Threshold}
\label{app:crthr}
\begin{figure}
    \centering
    \includegraphics[width=0.46\textwidth]{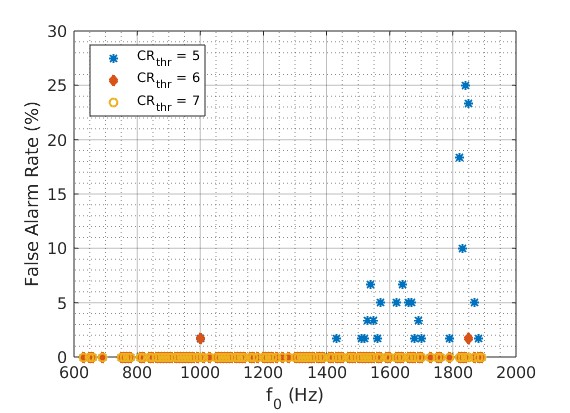}
    \caption{False-alarm rate as a function of frequency for different critical ratio thresholds $CR_{\mathrm{thr}}$. Most points are zero for higher thresholds (6 and 7), illustrating that the chosen $CR_{\rm thr}=7$ effectively suppresses false alarms across nearly all $f_0$ whereas $CR_{\rm thr}=5$ shows a non-negligible false-alarm fraction}
    \label{fig:crthr}
\end{figure}

The critical ratio threshold, $CR_{\mathrm{thr}}$, determines the balance between search sensitivity and false-alarm control in the candidate selection stage.  

A lower threshold increases the number of recovered signals but also raises the rate of false alarms, while a higher threshold reduces false alarms at the cost of missing weaker signals.  

To identify an optimal choice, we performed a set of injection tests on real O4a LIGO data at a fixed source distance of 1000~Mpc, using the same procedure as described in Section~\ref{sec:exp_sens}, analyzing the recovery fraction which is in this case the corresponding false-alarm rate for different values of $CR_{\mathrm{thr}}$.

Figure~\ref{fig:crthr} shows the false-alarm fraction versus frequency for different $CR_{\mathrm{thr}}$ values.  
As expected, the false-alarm rate decreases rapidly with increasing threshold.  
We found that values below $CR_{\mathrm{thr}} \approx 5$ lead to a significant rise in false coincidences, while values above 7 begin to substantially reduce the recovery efficiency for weaker signals.  

Based on these tests, we adopted $CR_{\mathrm{thr}} = 7$ as a compromise providing a stable false-alarm rate across the full frequency range while maintaining good detection efficiency.

\bibliographystyle{apsrev4-2}
\bibliography{references}

@ARTICLE{2015PASA3234L,
       author = {Lasky, Paul D.},
        title = {Gravitational Waves from Neutron Stars: A Review},
      journal = {Publications of the Astronomical Society of Australia},
         year = 2015,
        month = sep,
       volume = {32},
          eid = {e034},
        pages = {e034},
          doi = {10.1017/pasa.2015.35},
       eprint = {1508.06643},
}

@ARTICLE{2023LRR263R,
       author = {Riles, Keith},
        title = {Searches for continuous-wave gravitational radiation},
      journal = {Living Reviews in Relativity},
         year = 2023,
        month = dec,
       volume = {26},
       number = {1},
          eid = {3},
        pages = {3},
          doi = {10.1007/s41114-023-00044-3},
       eprint = {2206.06447},
}

@ARTICLE{2001A&A367525P,
       author = {Palomba, C.},
        title = {Gravitational radiation from young magnetars: Preliminary results},
      journal = {Astronomy and Astrophysics},
         year = 2001,
        month = feb,
       volume = {367},
        pages = {525-531},
          doi = {10.1051/0004-6361:20000452},
}

@ARTICLE{2002PhRvD66h4025C,
       author = {Cutler, Curt},
        title = {Gravitational waves from neutron stars with large toroidal B fields},
      journal = {Physical Review D},
         year = 2002,
        month = oct,
       volume = {66},
       number = {8},
          eid = {084025},
        pages = {084025},
          doi = {10.1103/PhysRevD.66.084025},
       eprint = {gr-qc/0206051},
}

@INPROCEEDINGS{2022ASSL465245D,
       author = {Dall'Osso, Simone and Stella, Luigi},
        title = {Millisecond Magnetars},
    booktitle = {Astrophysics and Space Science Library},
         year = 2022,
       editor = {Bhattacharyya, Sudip and Papitto, Alessandro and Bhattacharya, Dipankar},
       series = {Astrophysics and Space Science Library},
       volume = {465},
        month = jan,
        pages = {245-280},
          doi = {10.1007/978-3-030-85198-9_8},
       eprint = {2103.10878},
}

@ARTICLE{2021MNRAS5024680S,
       author = {Sur, Ankan and Haskell, Brynmor},
        title = {Gravitational waves from mountains in newly born millisecond magnetars},
      journal = {Monthly Notices of the Royal Astronomical Society},
         year = 2021,
        month = apr,
       volume = {502},
       number = {4},
        pages = {4680-4688},
          doi = {10.1093/mnras/stab307},
       eprint = {2010.15574},
}

@ARTICLE{2009ApJ701171C,
       author = {Corsi, Alessandra and Mészáros, Peter},
        title = {Gamma-ray Burst Afterglow Plateaus and Gravitational Waves: Multi-messenger Signature of a Millisecond Magnetar?},
      journal = {The Astrophysical Journal},
         year = 2009,
        month = sep,
       volume = {702},
       number = {2},
        pages = {1171-1178},
          doi = {10.1088/0004-637X/702/2/1171},
       eprint = {0907.2290},
}

@ARTICLE{2014PhRvD90d2002A,
       author = {Astone, Pia and Colla, Alberto and D'Antonio, Sabrina and Frasca, Sergio and Palomba, Cristiano},
        title = {Method for all-sky searches of continuous gravitational wave signals using the frequency-Hough transform},
      journal = {Physical Review D},
         year = 2014,
        month = aug,
       volume = {90},
       number = {4},
          eid = {042002},
        pages = {042002},
          doi = {10.1103/PhysRevD.90.042002},
       eprint = {1407.8333},
}

@ARTICLE{2018PhRvD98j2004M,
       author = {Miller, Andrew and Astone, Pia and D'Antonio, Sabrina and Frasca, Sergio and Intini, Giuseppe and La Rosa, Iuri and Leaci, Paola and Mastrogiovanni, Simone and Muciaccia, Federico and Palomba, Cristiano and Piccinni, Ornella J. and Singhal, Akshat and Whiting, Bernard F.},
        title = {Method to search for long duration gravitational wave transients from isolated neutron stars using the generalized frequency-Hough transform},
      journal = {Physical Review D},
         year = 2018,
        month = nov,
       volume = {98},
       number = {10},
          eid = {102004},
        pages = {102004},
          doi = {10.1103/PhysRevD.98.102004},
       eprint = {1810.09784},
}

@ARTICLE{2009MNRAS3981869D,
       author = {Dall'Osso, S. and Shore, S.N. and Stella, L.},
        title = {Early evolution of newly born magnetars with a strong toroidal field},
      journal = {Monthly Notices of the Royal Astronomical Society},
         year = 2009,
        month = oct,
       volume = {398},
       number = {4},
        pages = {1869-1885},
          doi = {10.1111/j.1365-2966.2008.14054.x},
       eprint = {0811.4311},
}

@ARTICLE{dallo15,
       author = {Dall'Osso, Simone and Giacomazzo, Bruno and Perna, Rosalba and Stella, Luigi},
        title = {Gravitational Waves from Massive Magnetars Formed in Binary Neutron Star Mergers},
      journal = {The Astrophysical Journal},
         year = 2015,
        month = jan,
       volume = {798},
       number = {1},
          eid = {25},
        pages = {25},
          doi = {10.1088/0004-637X/798/1/25},
       eprint = {1408.0013},
}

@ARTICLE{2018MNRAS4801353D,
       author = {Dall'Osso, S. and Stella, L. and Palomba, C.},
        title = {Neutron star bulk viscosity, 'spin-flip' and GW emission of newly born magnetars},
      journal = {Monthly Notices of the Royal Astronomical Society},
         year = 2018,
        month = oct,
       volume = {480},
       number = {1},
        pages = {1353-1362},
          doi = {10.1093/mnras/sty1706},
       eprint = {1806.11164},
}

@ARTICLE{2020MNRAS4944838L,
       author = {Lander, S.K. and Jones, D.I.},
        title = {Magnetar birth: rotation rates and gravitational-wave emission},
      journal = {Monthly Notices of the Royal Astronomical Society},
         year = 2020,
        month = jun,
       volume = {494},
       number = {4},
        pages = {4838-4847},
          doi = {10.1093/mnras/staa966},
       eprint = {1910.14336},
}

@ARTICLE{2019ApJ875160A,
       author = {Abbott, B.P. and others},
        title = {Search for Gravitational Waves from a Long-lived Remnant of the Binary Neutron Star Merger GW170817},
      journal = {The Astrophysical Journal},
         year = 2019,
        month = apr,
       volume = {875},
       number = {2},
          eid = {160},
        pages = {160},
          doi = {10.3847/1538-4357/ab0f3d},
       eprint = {1810.02581},
}

@ARTICLE{LATTIMER2016127,
title = {The equation of state of hot, dense matter and neutron stars},
journal = {Physics Reports},
volume = {621},
pages = {127-164},
year = {2016},
note = {Memorial Volume in Honor of Gerald E. Brown},
doi = {10.1016/j.physrep.2015.12.005},
author = {James M. Lattimer and Madappa Prakash},
}

@ARTICLE{2019MNRAS4871426B,
       author = {Beniamini, Paz and Hotokezaka, Kenta and van der Horst, Alexander and Kouveliotou, Chryssa},
        title = {Formation rates and evolution histories of magnetars},
      journal = {Monthly Notices of the Royal Astronomical Society},
         year = 2019,
        month = jul,
       volume = {487},
       number = {1},
        pages = {1426-1438},
          doi = {10.1093/mnras/stz1391},
       eprint = {1903.06718},
}

@ARTICLE{BSD_Piccinni_2019,
  author = {Piccinni, O.J. and Astone, P. and D’Antonio, S. and Frasca, S. and Intini, G. and Leaci, P. and Mastrogiovanni, S. and Miller, A. and Palomba, C. and Singhal, A.},
  title = {A new data analysis framework for the search of continuous gravitational wave signals},
  journal = {Classical and Quantum Gravity},
  volume = {36},
  number = {1},
  pages = {015008},
  year = {2018},
  doi = {10.1088/1361-6382/aaefb5},
}

@misc{Peaksens,
  author = {Cristiano Palomba},
  title = {{On the sensitivity of peakmap-based methods for the search of continuous gravitational wave signals}},
  howpublished = "\url{https://tds.virgo-gw.eu/?r=25138}",
  year = {2025}, 
  note = "Public Virgo note"
}

@misc{Itagaki2023_SN2023ixf,
  author = {Itagaki, K{\"o}ichi},
  title = {Discovery of SN 2023ixf in M101},
  howpublished = {\url{https://www.wis-tns.org/object/2023ixf/discovery-cert}},
  year = {2023},
  note = {Transient Name Server discovery certificate}
}

@ARTICLE{2023ApJ952L23K,
       author = {Kilpatrick, Charles D. and Foley, Ryan J. and Jacobson-Galán, Wynn V. and Piro, Anthony L. and Smartt, Stephen J. and Drout, Maria R. and Gagliano, Alexander and Gall, Christa and Hjorth, Jens and Jones, David O. and Mandel, Kaisey S. and Margutti, Raffaella and Ramirez-Ruiz, Enrico and Ransome, Conor L. and Villar, V. Ashley and Coulter, David A. and Gao, Hua and Matthews, David Jacob and Taggart, Kirsty and Zenati, Yossef},
        title = {SN 2023ixf in Messier 101: A Variable Red Supergiant as the Progenitor Candidate to a Type II Supernova},
      journal = {The Astrophysical Journal Letters},
         year = 2023,
        month = jul,
       volume = {952},
       number = {1},
          eid = {L23},
        pages = {L23},
          doi = {10.3847/2041-8213/ace4ca},
       eprint = {2306.04722},
}

@ARTICLE{2024A&A687L20F,
       author = {Ferrari, Lucía and Folatelli, Gastón and Ertini, Keila and Kuncarayakti, Hanindyo and Andrews, Jennifer E.},
        title = {Progenitor mass and ejecta asymmetry of supernova 2023ixf from nebular spectroscopy},
      journal = {Astronomy \& Astrophysics},
         year = 2024,
        month = jul,
       volume = {687},
          eid = {L20},
        pages = {L20},
          doi = {10.1051/0004-6361/202450440},
       eprint = {2406.00130},
}

@ARTICLE{2025arXiv250818079T,
       author = {The LIGO Scientific Collaboration and the Virgo Collaboration and the KAGRA Collaboration},
        title = {Open Data from LIGO, Virgo, and KAGRA through the First Part of the Fourth Observing Run},
      journal = {arXiv e-prints},
         year = 2025,
        month = aug,
          eid = {arXiv:2508.18079},
        pages = {arXiv:2508.18079},
          doi = {10.48550/arXiv.2508.18079},
       eprint = {2508.18079},
}

@ARTICLE{2020LRR233A,
       author = {Abbott, B.P. and others (LIGO Scientific Collaboration, Virgo Collaboration, KAGRA Collaboration)},
      title = {Prospects for observing and localizing gravitational-wave transients with Advanced LIGO, Advanced Virgo and KAGRA},
      journal = {Living Reviews in Relativity},
         year = 2020,
        month = dec,
       volume = {23},
       number = {1},
          eid = {3},
        pages = {3},
          doi = {10.1007/s41114-020-00026-9},
}

@ARTICLE{Maggiore2020_ET,
  author = {Maggiore, M. and others},
  title = {Science case for the Einstein Telescope},
  journal = {Journal of Cosmology and Astroparticle Physics},
  number = {03},
  pages = {050},
  year = {2020},
  doi = {10.1088/1475-7516/2020/03/050},
}

@ARTICLE{Evans2021_CE,
  author = {Evans, E.D. and others},
  title = {Gravitational-wave physics with Cosmic Explorer: Limits to low frequency sensitivity, and parameter estimation across the mass range},
  journal = {Physical Review D},
  volume = {103},
  pages = {122004},
  year = {2021},
  doi = {10.1103/PhysRevD.103.122004},
}

@ARTICLE{2004PhRvD70h2001K,
       author = {Krishnan, Badri and Sintes, Alicia M. and Papa, Maria Alessandra and Schutz, Bernard F. and Frasca, Sergio and Palomba, Cristiano},
        title = {Hough transform search for continuous gravitational waves},
      journal = {Physical Review D},
         year = 2004,
        month = oct,
       volume = {70},
       number = {8},
          eid = {082001},
        pages = {082001},
          doi = {10.1103/PhysRevD.70.082001},
       eprint = {gr-qc/0407001},
}

@ARTICLE{2019PhRvD99j4067O,
       author = {Oliver, Miquel and Keitel, David and Sintes, Alicia M.},
        title = {Adaptive transient Hough method for long-duration gravitational wave transients},
      journal = {Physical Review D},
         year = 2019,
        month = may,
       volume = {99},
       number = {10},
          eid = {104067},
        pages = {104067},
          doi = {10.1103/PhysRevD.99.104067},
       eprint = {1901.01820},
}

@ARTICLE{2024PhRvD110j3047A,
       author = {Attadio, Francesca and Ricca, Leonardo and Serra, Marco and Palomba, Cristiano and Astone, Pia and Dall'Osso, Simone and Dal Pra, Stefano and D'Antonio, Sabrina and Di Giovanni, Matteo and D'Onofrio, Luca and Leaci, Paola and Muciaccia, Federico and Pierini, Lorenzo and Safai Tehrani, Francesco},
        title = {Neural network method to search for long transient gravitational waves},
      journal = {Physical Review D},
         year = 2024,
        month = nov,
       volume = {110},
       number = {10},
          eid = {103047},
        pages = {103047},
          doi = {10.1103/PhysRevD.110.103047},
       eprint = {2407.02391},
}

@ARTICLE{2012ApJ76163P,
       author = {Piro, Anthony L. and Thrane, Eric},
        title = {Gravitational Waves from Fallback Accretion onto Neutron Stars},
      journal = {The Astrophysical Journal},
         year = 2012,
        month = dec,
       volume = {761},
       number = {1},
          eid = {63},
        pages = {63},
          doi = {10.1088/0004-637X/761/1/63},
       eprint = {1207.3805},
}

@ARTICLE{2011PhRvD83j4014C,
       author = {Corsi, Alessandra and Owen, Benjamin J.},
        title = {Maximum gravitational-wave energy emissible in magnetar flares},
      journal = {Physical Review D},
         year = 2011,
        month = may,
       volume = {83},
       number = {10},
          eid = {104014},
        pages = {104014},
          doi = {10.1103/PhysRevD.83.104014},
       eprint = {1102.3421},
}

@ARTICLE{2015PhRvD91f3007H,
       author = {Hamil, O. and Stone, J.R. and Urbanec, M. and Urbancová, G.},
        title = {Braking index of isolated pulsars},
      journal = {Physical Review D},
         year = 2015,
        month = mar,
       volume = {91},
       number = {6},
          eid = {063007},
        pages = {063007},
          doi = {10.1103/PhysRevD.91.063007},
       eprint = {1608.01383},
}

@ARTICLE{2011GReGr43409A,
       author = {Andersson, N. and Ferrari, V. and Jones, D.I. and Kokkotas, K.D. and Krishnan, B. and Read, J.S. and Rezzolla, L. and Zink, B.},
        title = {Gravitational waves from neutron stars: promises and challenges},
      journal = {General Relativity and Gravitation},
         year = 2011,
        month = feb,
       volume = {43},
       number = {2},
        pages = {409-436},
          doi = {10.1007/s10714-010-1059-4},
       eprint = {0912.0384},
}

@ARTICLE{Rosa:2021ptb,
    author = {Rosa, Iuri La and Astone, Pia and D'Antonio, Sabrina and Frasca, Sergio and Leaci, Paola and Miller, Andrew Lawrence and Palomba, Cristiano and Piccinni, Ornella Juliana and Pierini, Lorenzo and Regimbau, Tania},
    title = {Continuous Gravitational-Wave Data Analysis with General Purpose Computing on Graphic Processing Units},
    doi = {10.3390/universe7070218},
    journal = {Universe},
    volume = {7},
    number = {7},
    pages = {218},
    year = {2021}
}

@ARTICLE{2015CQGra32b4001A,
       author = {Acernese, F. and others},
        title = {Advanced Virgo: a second-generation interferometric gravitational wave detector},
      journal = {Classical and Quantum Gravity},
         year = 2015,
        month = jan,
       volume = {32},
       number = {2},
          eid = {024001},
        pages = {024001},
          doi = {10.1088/0264-9381/32/2/024001},
       eprint = {1408.3978},
}

@ARTICLE{2015CQGra32g4001L,
       author = {Aasi, J. and others},
        title = {Advanced LIGO},
      journal = {Classical and Quantum Gravity},
         year = 2015,
        month = apr,
       volume = {32},
       number = {7},
          eid = {074001},
        pages = {074001},
          doi = {10.1088/0264-9381/32/7/074001},
       eprint = {1411.4547},
}

@ARTICLE{2021PTEP2021eA102A,
       author = {Akutsu, T. and others},
        title = {Overview of KAGRA: Calibration, detector characterization, physical environmental monitors, and the geophysics interferometer},
      journal = {Progress of Theoretical and Experimental Physics},
         year = 2021,
        month = may,
       volume = {2021},
       number = {5},
          eid = {05A102},
        pages = {05A102},
          doi = {10.1093/ptep/ptab018},
       eprint = {2009.09305},
}

@ARTICLE{2002CQGra191377W,
       author = {Willke, B. and others},
        title = {The GEO 600 gravitational wave detector},
      journal = {Classical and Quantum Gravity},
         year = 2002,
        month = apr,
       volume = {19},
       number = {7},
        pages = {1377-1387},
          doi = {10.1088/0264-9381/19/7/321},
}

@book{Maggiore:2007ulw,
    author = "Maggiore, Michele",
    title = "{Gravitational Waves. Vol. 1: Theory and Experiments}",
    doi = "10.1093/acprof:oso/9780198570745.001.0001",
    isbn = "978-0-19-171766-6, 978-0-19-852074-0",
    publisher = "Oxford University Press",
    year = "2007"
}

@ARTICLE{1998PhRvD58f3001J,
       author = {Jaranowski, Piotr and Królak, Andrzej and Schutz, Bernard F.},
        title = {Data analysis of gravitational-wave signals from spinning neutron stars: The signal and its detection},
      journal = {Physical Review D},
         year = 1998,
        month = sep,
       volume = {58},
       number = {6},
          eid = {063001},
        pages = {063001},
          doi = {10.1103/PhysRevD.58.063001},
       eprint = {gr-qc/9804014},
}

@book{Shapiro:1983du,
    author = "Shapiro, S. L. and Teukolsky, S. A.",
    title = "{Black holes, white dwarfs, and neutron stars: The physics of compact objects}",
    doi = "10.1002/9783527617661",
    isbn = "978-0-471-87316-7, 978-3-527-61766-1",
    publisher = "Wiley",
    year = "1983"
}

@misc{GWOSC,
  author = {Gravitational Wave Open Science Center},
  title = {GWOSC},
  year = {2025},
  note = {Available at https://gwosc.org/O4/O4a/}
}

@ARTICLE{2019PhRvD.100f2005M,
       author = {{Miller}, Andrew L. and {Astone}, Pia and {D'Antonio}, Sabrina and {Frasca}, Sergio and {Intini}, Giuseppe and {La Rosa}, Iuri and {Leaci}, Paola and {Mastrogiovanni}, Simone and {Muciaccia}, Federico and {Mitidis}, Andonis and {Palomba}, Cristiano and {Piccinni}, Ornella J. and {Singhal}, Akshat and {Whiting}, Bernard F. and {Rei}, Luca},
        title = "{How effective is machine learning to detect long transient gravitational waves from neutron stars in a real search?}",
      journal = {\prd},
         year = 2019,
        month = sep,
       volume = {100},
       number = {6},
          eid = {062005},
        pages = {062005},
          doi = {10.1103/PhysRevD.100.062005},
archivePrefix = {arXiv},
       eprint = {1909.02262},
 primaryClass = {astro-ph.IM},
       adsurl = {https://ui.adsabs.harvard.edu/abs/2019PhRvD.100f2005M}
}

@ARTICLE{2011PhRvD84b3007P,
       author = {{Prix}, R. and {Giampanis}, S. and {Messenger}, C.},
        title = "{Search method for long-duration gravitational-wave transients from neutron stars}",
      journal = {\prd},
         year = 2011,
        month = jul,
       volume = {84},
       number = {2},
          eid = {023007},
        pages = {023007},
          doi = {10.1103/PhysRevD.84.023007},
archivePrefix = {arXiv},
       eprint = {1104.1704},
 primaryClass = {gr-qc},
       adsurl = {https://ui.adsabs.harvard.edu/abs/2011PhRvD..84b3007P}
}

@ARTICLE{2008PhRvD78d4031K,
       author = {{Knispel}, Benjamin and {Allen}, Bruce},
        title = "{Blandford's argument: The strongest continuous gravitational wave signal}",
      journal = {\prd},
         year = 2008,
        month = aug,
       volume = {78},
       number = {4},
          eid = {044031},
        pages = {044031},
          doi = {10.1103/PhysRevD.78.044031},
archivePrefix = {arXiv},
       eprint = {0804.3075},
 primaryClass = {gr-qc},
       adsurl = {https://ui.adsabs.harvard.edu/abs/2008PhRvD..78d4031K}
}

@ARTICLE{2025PhRvD.111f2002C,
       author = {{Capote}, E. and others},
        title = "{Advanced LIGO detector performance in the fourth observing run}",
      journal = {\prd},
         year = 2025,
        month = mar,
       volume = {111},
       number = {6},
          eid = {062002},
        pages = {062002},
          doi = {10.1103/PhysRevD.111.062002},
archivePrefix = {arXiv},
       eprint = {2411.14607},
 primaryClass = {gr-qc},
       adsurl = {https://ui.adsabs.harvard.edu/abs/2025PhRvD.111f2002C}
}

@ARTICLE{1995ApJ442259L,
       author = {{Lai}, Dong and {Shapiro}, Stuart L.},
        title = "{Gravitational Radiation from Rapidly Rotating Nascent Neutron Stars}",
      journal = {\apj},
         year = 1995,
        month = mar,
       volume = {442},
        pages = {259},
          doi = {10.1086/175438},
archivePrefix = {arXiv},
       eprint = {astro-ph/9408053},
 primaryClass = {astro-ph},
       adsurl = {https://ui.adsabs.harvard.edu/abs/1995ApJ...442..259L}
}

@ARTICLE{2000MNRAS.319902U,
       author = {{Ushomirsky}, Greg and {Cutler}, Curt and {Bildsten}, Lars},
        title = "{Deformations of accreting neutron star crusts and gravitational wave emission}",
      journal = {Monthly Notices of the Royal Astronomical Society},
         year = 2000,
        month = dec,
       volume = {319},
       number = {3},
        pages = {902-932},
          doi = {10.1046/j.1365-8711.2000.03938.x},
archivePrefix = {arXiv},
       eprint = {astro-ph/0001136},
 primaryClass = {astro-ph},
       adsurl = {https://ui.adsabs.harvard.edu/abs/2000MNRAS.319..902U}
}

@ARTICLE{2008MNRAS.385531H,
       author = {{Haskell}, B. and {Samuelsson}, L. and {Glampedakis}, K. and {Andersson}, N.},
        title = "{Modelling magnetically deformed neutron stars}",
      journal = {Monthly Notices of the Royal Astronomical Society},
         year = 2008,
        month = mar,
       volume = {385},
       number = {1},
        pages = {531-542},
          doi = {10.1111/j.1365-2966.2008.12861.x},
archivePrefix = {arXiv},
       eprint = {0705.1780},
 primaryClass = {astro-ph},
       adsurl = {https://ui.adsabs.harvard.edu/abs/2008MNRAS.385..531H}
}

@ARTICLE{2005CQGra22S1197A,
       author = {{Astone}, P. and {Frasca}, S. and {Palomba}, C.},
        title = "{The short FFT database and the peak map for the hierarchical search of periodic sources}",
      journal = {Classical and Quantum Gravity},
         year = 2005,
        month = sep,
       volume = {22},
       number = {18},
        pages = {S1197-S1210},
          doi = {10.1088/0264-9381/22/18/S34},
       adsurl = {https://ui.adsabs.harvard.edu/abs/2005CQGra..22S1197A}
}

@ARTICLE{2020CQGra37v5008S,
       author = {{Sun}, Ling and {Goetz}, Evan and {Kissel}, Jeffrey S. and {Betzwieser}, Joseph and {Karki}, Sudarshan and {Viets}, Aaron and {Wade}, Madeline and {Bhattacharjee}, Dripta and {Bossilkov}, Vladimir and {Covas}, Pep B. and {Datrier}, Laurence E.~H. and {Gray}, Rachel and {Kandhasamy}, Shivaraj and {Lecoeuche}, Yannick K. and {Mendell}, Gregory and {Mistry}, Timesh and {Payne}, Ethan and {Savage}, Richard L. and {Weinstein}, Alan J. and {Aston}, Stuart and {Buikema}, Aaron and {Cahillane}, Craig and {Driggers}, Jenne C. and {Dwyer}, Sheila E. and {Kumar}, Rahul and {Urban}, Alexander},
        title = "{Characterization of systematic error in Advanced LIGO calibration}",
      journal = {Classical and Quantum Gravity},
         year = 2020,
        month = nov,
       volume = {37},
       number = {22},
          eid = {225008},
        pages = {225008},
          doi = {10.1088/1361-6382/abb14e},
archivePrefix = {arXiv},
       eprint = {2005.02531},
 primaryClass = {astro-ph.IM},
       adsurl = {https://ui.adsabs.harvard.edu/abs/2020CQGra..37v5008S}
}

@ARTICLE{2022CQGra39d5006A,
       author = {{Acernese}, F. and others},
        title = "{Calibration of advanced Virgo and reconstruction of the detector strain h(t) during the observing run O3}",
      journal = {Classical and Quantum Gravity},
         year = 2022,
        month = feb,
       volume = {39},
       number = {4},
          eid = {045006},
        pages = {045006},
          doi = {10.1088/1361-6382/ac3c8e},
archivePrefix = {arXiv},
       eprint = {2107.03294},
 primaryClass = {gr-qc},
       adsurl = {https://ui.adsabs.harvard.edu/abs/2022CQGra..39d5006A}
}

@ARTICLE{2023APh15302880W,
       author = {{Wette}, Karl},
        title = "{Searches for continuous gravitational waves from neutron stars: A twenty-year retrospective}",
      journal = {Astroparticle Physics},
         year = 2023,
        month = nov,
       volume = {153},
          eid = {102880},
        pages = {102880},
          doi = {10.1016/j.astropartphys.2023.102880},
archivePrefix = {arXiv},
       eprint = {2305.07106},
 primaryClass = {gr-qc},
       adsurl = {https://ui.adsabs.harvard.edu/abs/2023APh...15302880W}
}

@ARTICLE{2022ApJ9351A,
       author = {{Abbott}, R. and others},
        title = "{Searches for Gravitational Waves from Known Pulsars at Two Harmonics in the Second and Third LIGO-Virgo Observing Runs}",
      journal = {\apj},
         year = 2022,
        month = aug,
       volume = {935},
       number = {1},
          eid = {1},
        pages = {1},
          doi = {10.3847/1538-4357/ac6acf},
archivePrefix = {arXiv},
       eprint = {2111.13106},
 primaryClass = {astro-ph.HE},
       adsurl = {https://ui.adsabs.harvard.edu/abs/2022ApJ...935....1A}
}

@ARTICLE{2019EPJA5580R,
       author = {{Raithel}, Carolyn A.},
        title = "{Constraints on the neutron star equation of state from GW170817}",
      journal = {European Physical Journal A},
         year = 2019,
        month = may,
       volume = {55},
       number = {5},
          eid = {80},
        pages = {80},
          doi = {10.1140/epja/i2019-12759-5},
archivePrefix = {arXiv},
       eprint = {1904.10002},
 primaryClass = {astro-ph.HE},
       adsurl = {https://ui.adsabs.harvard.edu/abs/2019EPJA...55...80R}
}

@ARTICLE{2022NatRP4237Y,
       author = {{Yunes}, Nicol{\'a}s and {Miller}, M. Coleman and {Yagi}, Kent},
        title = "{Gravitational-wave and X-ray probes of the neutron star equation of state}",
      journal = {Nature Reviews Physics},
         year = 2022,
        month = feb,
       volume = {4},
       number = {4},
        pages = {237-246},
          doi = {10.1038/s42254-022-00420-y},
archivePrefix = {arXiv},
       eprint = {2202.04117},
 primaryClass = {gr-qc},
       adsurl = {https://ui.adsabs.harvard.edu/abs/2022NatRP...4..237Y}
}

@ARTICLE{2005CQGra22S1255P,
       author = {{Palomba}, Cristiano and {Astone}, Pia and {Frasca}, Sergio},
        title = "{Adaptive Hough transform for the search of periodic sources}",
      journal = {Classical and Quantum Gravity},
         year = 2005,
        month = sep,
       volume = {22},
       number = {18},
        pages = {S1255-S1264},
          doi = {10.1088/0264-9381/22/18/S39},
       adsurl = {https://ui.adsabs.harvard.edu/abs/2005CQGra..22S1255P}
}

@ARTICLE{2019PhRvD.100b4034B,
       author = {{Banagiri}, Sharan and {Sun}, Ling and {Coughlin}, Michael W. and {Melatos}, Andrew},
        title = "{Search strategies for long gravitational-wave transients: Hidden Markov model tracking and seedless clustering}",
      journal = {\prd},
         year = 2019,
        month = jul,
       volume = {100},
       number = {2},
          eid = {024034},
        pages = {024034},
          doi = {10.1103/PhysRevD.100.024034},
archivePrefix = {arXiv},
       eprint = {1903.02638},
 primaryClass = {astro-ph.IM},
       adsurl = {https://ui.adsabs.harvard.edu/abs/2019PhRvD.100b4034B}
}

@ARTICLE{2018CQGra35t5003K,
       author = {{Keitel}, David and {Ashton}, Gregory},
        title = "{Faster search for long gravitational-wave transients: GPU implementation of the transient \textbackslashnewcommand\{\textbackslashF\}F\textbackslashboldsymbol\{ \textbackslashF\} -statistic}",
      journal = {Classical and Quantum Gravity},
         year = 2018,
        month = oct,
       volume = {35},
       number = {20},
          eid = {205003},
        pages = {205003},
          doi = {10.1088/1361-6382/aade34},
archivePrefix = {arXiv},
       eprint = {1805.05652},
 primaryClass = {astro-ph.IM},
       adsurl = {https://ui.adsabs.harvard.edu/abs/2018CQGra..35t5003K}
}

@ARTICLE{2019PhRvD.100f4058K,
       author = {{Keitel}, David and {Woan}, Graham and {Pitkin}, Matthew and {Schumacher}, Courtney and {Pearlstone}, Brynley and {Riles}, Keith and {Lyne}, Andrew G. and {Palfreyman}, Jim and {Stappers}, Benjamin and {Weltevrede}, Patrick},
        title = "{First search for long-duration transient gravitational waves after glitches in the Vela and Crab pulsars}",
      journal = {\prd},
         year = 2019,
        month = sep,
       volume = {100},
       number = {6},
          eid = {064058},
        pages = {064058},
          doi = {10.1103/PhysRevD.100.064058},
archivePrefix = {arXiv},
       eprint = {1907.04717},
 primaryClass = {gr-qc},
       adsurl = {https://ui.adsabs.harvard.edu/abs/2019PhRvD.100f4058K}
}

@ARTICLE{2019PhRvD99l3003S,
       author = {{Sun}, Ling and {Melatos}, Andrew},
        title = "{Application of hidden Markov model tracking to the search for long-duration transient gravitational waves from the remnant of the binary neutron star merger GW170817}",
      journal = {\prd},
         year = 2019,
        month = jun,
       volume = {99},
       number = {12},
          eid = {123003},
        pages = {123003},
          doi = {10.1103/PhysRevD.99.123003},
archivePrefix = {arXiv},
       eprint = {1810.03577},
 primaryClass = {astro-ph.IM},
       adsurl = {https://ui.adsabs.harvard.edu/abs/2019PhRvD..99l3003S}
}

@ARTICLE{2016PhRvD93h4024K,
       author = {{Keitel}, David},
        title = "{Robust semicoherent searches for continuous gravitational waves with noise and signal models including hours to days long transients}",
      journal = {\prd},
         year = 2016,
        month = apr,
       volume = {93},
       number = {8},
          eid = {084024},
        pages = {084024},
          doi = {10.1103/PhysRevD.93.084024},
archivePrefix = {arXiv},
       eprint = {1509.02398},
 primaryClass = {gr-qc},
       adsurl = {https://ui.adsabs.harvard.edu/abs/2016PhRvD..93h4024K}
}

@ARTICLE{2024PhRvD.109l3516A,
       author = {{Alestas}, George and {Morr{\'a}s}, Gonzalo and {Yamamoto}, Takahiro S. and {Garc{\'\i}a-Bellido}, Juan and {Kuroyanagi}, Sachiko and {Nesseris}, Savvas},
        title = "{Applying the Viterbi algorithm to planetary-mass black hole searches}",
      journal = {\prd},
         year = 2024,
        month = jun,
       volume = {109},
       number = {12},
          eid = {123516},
        pages = {123516},
          doi = {10.1103/PhysRevD.109.123516},
archivePrefix = {arXiv},
       eprint = {2401.02314},
 primaryClass = {astro-ph.CO},
       adsurl = {https://ui.adsabs.harvard.edu/abs/2024PhRvD.109l3516A}
}

@ARTICLE{2025PhRvD.111d3019A,
       author = {{Andr{\'e}s-Carcasona}, M. and {Piccinni}, O.~J. and {Mart{\'\i}nez}, M. and {Mir}, Ll. M.},
        title = "{New approach to search for long transient gravitational waves from inspiraling compact binary systems}",
      journal = {\prd},
         year = 2025,
        month = feb,
       volume = {111},
       number = {4},
          eid = {043019},
        pages = {043019},
          doi = {10.1103/PhysRevD.111.043019},
archivePrefix = {arXiv},
       eprint = {2411.04498},
 primaryClass = {gr-qc},
       adsurl = {https://ui.adsabs.harvard.edu/abs/2025PhRvD.111d3019A}
}

@misc{Pierini2025GFHv2,
  author       = {Pierini, L.},
  title        = {Generalized Frequency Hough (GFH-v2)},
  year         = {2025},
  howpublished = {New core function developed under ICSC (2.0)},
  doi          = {10.15161/oar.it/35swc-fes27},
  url          = {https://doi.org/10.15161/oar.it/35swc-fes27}
}

@misc{Astone2025SFDB,
  author = {Astone, P. and D'Antonio, S.},
  title = {Creation of the Short FFT data base (SFDB)},
  year = {2025},
  howpublished = {oar.it},
  doi = {10.15161/oar.it/tqwgf-nq452},
  url = {https://doi.org/10.15161/oar.it/tqwgf-nq452}
}

@ARTICLE{2022ApJ934125X,
       author = {{Xie}, Lang and {Wei}, Da-Ming and {Wang}, Yun and {Jin}, Zhi-Ping},
        title = "{Constraining the Ellipticity of the Newborn Magnetar with the Observational Data of Long Gamma-Ray Bursts}",
      journal = {Astrophysical Journal},
         year = 2022,
        month = aug,
       volume = {934},
       number = {2},
          eid = {125},
        pages = {125},
          doi = {10.3847/1538-4357/ac7c13},
archivePrefix = {arXiv},
       eprint = {2206.12874},
 primaryClass = {astro-ph.HE},
       adsurl = {https://ui.adsabs.harvard.edu/abs/2022ApJ...934..125X}
}

@ARTICLE{2023ApJ952156S,
       author = {{Song}, Cui-Ying and {Liu}, Tong},
        title = "{Long-duration Gamma-Ray Burst Progenitors and Magnetar Formation}",
      journal = {Astrophysical Journal},
         year = 2023,
        month = aug,
       volume = {952},
       number = {2},
          eid = {156},
        pages = {156},
          doi = {10.3847/1538-4357/acd6ee},
archivePrefix = {arXiv},
       eprint = {2301.05401},
 primaryClass = {astro-ph.HE},
       adsurl = {https://ui.adsabs.harvard.edu/abs/2023ApJ...952..156S}
}

@ARTICLE{2007Ap&SS308119D,
       author = {{Dall'Osso}, S. and {Stella}, L.},
        title = "{Newborn magnetars as sources of gravitational radiation: constraints from high energy observations of magnetar candidates}",
      journal = {Astrophysics and Space Science},
         year = 2007,
        month = apr,
       volume = {308},
       number = {1-4},
        pages = {119-124},
          doi = {10.1007/s10509-007-9323-0},
archivePrefix = {arXiv},
       eprint = {astro-ph/0702075},
 primaryClass = {astro-ph},
       adsurl = {https://ui.adsabs.harvard.edu/abs/2007Ap&SS.308..119D}
}

@misc{GWOSC_O4_SpecLines,
  author       = {{Gravitational Wave Open Science Center}},
  title        = {O4 Instrumental Lines},
  year         = {2024},
  url          = {https://gwosc.org/O4/o4speclines/},
}
\end{document}